\documentclass[acmsmall]{acmart}
\usepackage{tcolorbox}
\usepackage{enumitem}
\usepackage{multirow}
\usepackage{subcaption}
\usepackage{xcolor}
\definecolor{mydarkgreen}{rgb}{0.0, 0.5, 0.0}

\AtBeginDocument{%
  }

\setcopyright{acmlicensed}
\copyrightyear{2025}
\acmYear{2025}
\acmDOI{XXXXXXX.XXXXXXX}

\acmJournal{JACM}
\acmVolume{37}
\acmNumber{4}
\acmArticle{111}
\acmMonth{8}

\begin{document}

\title{Smart Cuts: Enhance Active Learning for Vulnerability Detection by Pruning Hard-to-Learn Data}


\author{Xiang Lan}
\affiliation{%
  \institution{North Carolina State University}
  \city{Raleigh}
  \country{USA}}
\email{xlan4@ncsu.edu}

\author{Tim Menzies}
\affiliation{%
  \institution{North Carolina State University}
  \city{Raleigh}
  \country{USA}}
\email{timm@ieee.org}

\author{Bowen Xu}
\affiliation{%
  \institution{North Carolina State University}
  \city{Raleigh}
  \country{USA}}
\email{bxu22@ncsu.edu}

\renewcommand{\shortauthors}{Lan et al.}

\begin{abstract}
  Vulnerability detection is crucial for identifying security weaknesses in software systems. However, training effective machine learning models for this task is often constrained by the high cost and expertise required for data annotation. Active learning is a promising approach to mitigate this challenge by intelligently selecting the most informative data points for labeling. This paper proposes a novel method to significantly enhance the active learning process by using dataset maps. Our approach systematically identifies samples that are \textit{hard-to-learn} for a model and integrates this information to create a more sophisticated sample selection strategy. Unlike traditional active learning methods that focus primarily on model uncertainty, our strategy enriches the selection process by considering learning difficulty, allowing the active learner to more effectively pinpoint truly informative examples. The experimental results show that our approach can improve F1 score over random selection by 61.54\% (DeepGini) and 45.91\% (K-Means) and outperforms standard active learning by 8.23\% (DeepGini) and 32.65\% (K-Means) for CodeBERT on the Big-Vul dataset, demonstrating the effectiveness of integrating dataset maps for optimizing sample selection in vulnerability detection. Furthermore, our approach also enhances model robustness, improves sample selection by filtering hard-to-learn data, and stabilizes active learning performance across iterations. By analyzing the characteristics of these outliers, we provide insights for future improvements in dataset construction, making vulnerability detection more reliable and cost-effective.
\end{abstract}

\ccsdesc[500]{Security and privacy~Software and application security}

\keywords{vulnerability detection,large language models}


\maketitle

\section{Introduction}
Vulnerability detection aims to identify weaknesses in software systems to prevent potential security attacks and privacy leakage. With the expanding capabilities of code language models (LMs), there is a surge in research utilizing them to autonomously identify security vulnerabilities~\cite{chakraborty2021deep,fu2022linevul,chen2023diversevul,steenhoek2023empirical}. This trend has created a pressing demand for high-quality vulnerability detection datasets to train and evaluate these powerful models~\cite{fan2020ac,chakraborty2021deep,nikitopoulos2021crossvul,bhandari2021cvefixes,chen2023diversevul}. However, a primary challenge in this domain is the significant cost and effort required to accurately label a significant amount of vulnerability data, a process that demands extensive security expert knowledge. 

Active learning is such a technique to address this issue that uses an acquisition function to select the most informative samples from a large pool of unlabeled data—for instance, those samples the model is most uncertain about. These newly labeled samples are then added to the training set to retrain and improve the model. This cycle is repeated, allowing the model to achieve high performance with significantly fewer labels than would be required by random sampling. This paradigm has been well-studied and successfully applied in other domains with high annotation costs, such as computer vision and natural language processing~\cite{yang2003automatically,li2013adaptive,zhang2022survey}. Inspired by this success, some recent works have begun to adapt active learning for vulnerability detection, seeking to build robust models while minimizing reliance on expert annotation~\cite{yu2019improving,hu2021towards,hu2024active}.

While active learning provides a framework for efficient data selection, our preliminary experiments (detailed in Section~\ref{sec:motivation}) reveal a counter-intuitive finding that underscores the complexity of this task. We demonstrated that training a detection model on a partial (e.g., 70\%) subset of data can even outperform a model trained on the entire dataset. This crucial observation implies that not all data necessarily contributes to the model's performance; some samples may not facilitate learning effectively. This raises a critical question for the active learning process: 

\begin{quote}
{\em How can we refine the sample selection strategy by detecting instances that are more likely to be detrimental?}\end{quote}

We hypothesize that by identifying these instances and mitigating their influence within the active learning loop, we can significantly enhance the effectiveness of the model training process.

To validate our hypothesis, we develop a training dynamic-based method to
categorize samples based on the difficulties they are learned by the model and then adopt it into an active learning framework.
By leveraging the insights of training dynamics, we systematically identify samples that do not substantially contribute to the model's performance improvement, referred to as \textit{hard-to-learn} samples. Next, we integrate these identified hard-to-learn samples into the active learning framework to evaluate their impact on model improvement. Specifically, we examine whether prioritizing these challenging samples in the active learning process can enhance its efficiency in selecting informative instances for annotation. Unlike conventional active learning strategies that focus on uncertainty-based or diversity-based sampling~\cite{li2012sample,hu2021towards,hu2024active}. By iteratively refining the training set in this manner, our method aims to enhance the standard active learning framework. The primary goal is to accelerate the model's learning curve, enabling it to achieve high performance more rapidly over the course of the active learning iterations. Concurrently, this approach fosters a more stable and consistent training dynamic. The ultimate outcome is an improved active learning cycle that produces more robust and effective vulnerability detection models compared to conventional strategies.

To demonstrate the effectiveness of our proposed method, we conduct a series of experiments. First, we collect dataset maps based on training dynamics to identify hard-to-learn data. Second, we incorporate the insights of learning difficulty into our active learning framework and refine the acquisition function. Third, we examine the generalizability of our approach by extending the experiments to multiple models and datasets.
In summary, this paper makes the following contributions to address the critical data challenge of vulnerability detection:

\begin{itemize}
    \item We propose a novel method for vulnerability detection from a new angle, i.e., the insights embedded in the model training dynamics.
    \item We incorporate these hard-to-learn data instances identified by our approach into active learning framework. We conducted a comprehensive evaluation across multiple datasets and models.
    \item We interpret common features of hard-to-learn data, providing actionable insights to advance vulnerability detection.
    \item We release datasets, scripts, and raw results used for conducting the study in our replication package \cite{badseeds} to facilitate future research.
\end{itemize}

\section{Problem and Motivation}\label{sec:motivation}
A widely held belief in software engineering is that more data leads to better models. For example:

\begin{itemize}
  \item ``\textit{..as long as it is large; the resulting prediction performance is likely to be boosted more by the size of the sample than it is hindered by any bias polarity that may exist}''~\cite{rahman2013sample}. 
  \item ``\textit{It is natural to think that a closer previous release has more similar characteristics and thus can help to train a more accurate defect prediction model. It is also natural to think that accumulating multiple releases can be beneficial because it represents the variability of a project}''
  ~\cite{amasaki2020cross}.
  \item ``\textit{Long-term JIT models should be trained using a cache of plenty of changes}''~\cite{mcintosh2018fix}.
\end{itemize}

Similar to the defect prediction problem, vulnerability detection is also commonly formulated as a binary classification problem. Vulnerability detection aims to determine whether a given code $c$ is vulnerable or not, i.e., \[f(c) \to \mathcal{Y} \quad \text{while} \quad \mathcal{Y} = \{\textit{vulnerable}, \textit{non-Vulnerable} \}\]

Recently, large language model (LLM)-based solutions have shown their efficiency in  solving the problem of  vulnerability detection~\cite{akuthota2023vulnerability,chakraborty2024revisiting,chakraborty2021deep,ding2024vulnerability,steenhoek2023empirical,lin2020software}.
As a technique with a data-hungry nature, the effectiveness of LLM also generally follows \textit{more data leads to better performance}.
Considering the above, we are interested in answering the question,

\begin{quote}
\textit{Is the common sense of ``the more data the better'' hold in the existing vulnerability detection context?}
\end{quote}

\begin{figure}[h]
    \centering
    \resizebox{.8\columnwidth}{!}{%
    \includegraphics{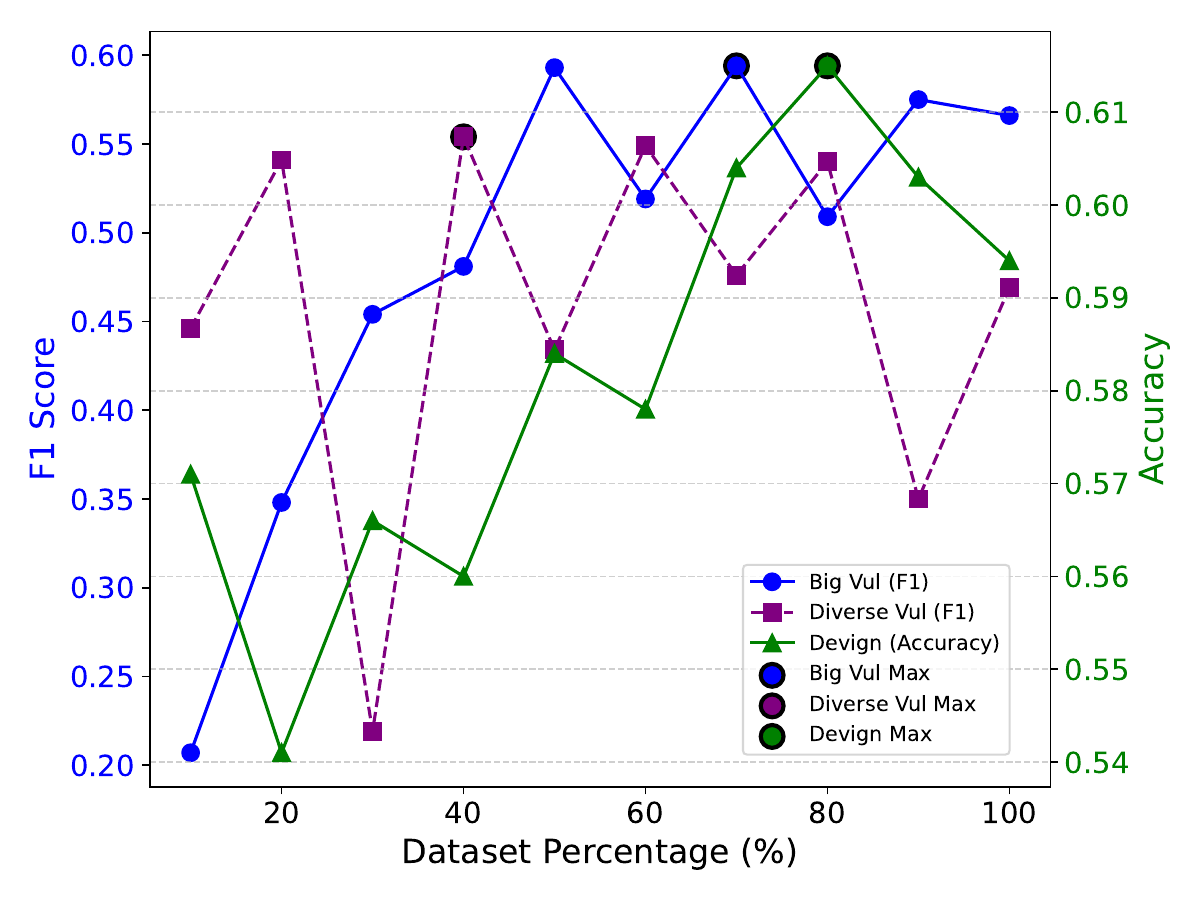}}
    \caption{Comparison of F1 Score and Accuracy across Different Dataset Ratio with Highlighted Max Values}
    \label{fig:f1_accuracy}
\end{figure}

To answer the question, we conduct a preliminary experiment on three commonly used vulnerability detection datasets, Big-Vul~\cite{fan2020ac}, Devign~\cite{zhou2019devign} and DiverseVul~\cite{chen2023diversevul}. For each dataset, we fine-tune a widely used LLM (i.e., CodeBERT~\cite{feng2020codebert}) based on 10 versions of datasets with gradually incremental sizes - starting from 10\% to 100\% with 10\% interval increase (i.e., 10\%, 20\%, ..., 100\%) — and evaluate the model's performance.
For each of these 10 data subsets, the model is trained for a maximum of 10 epochs, and we employ a standard early stopping mechanism to prevent overfitting.
As shown in Figure \ref{fig:f1_accuracy}, we find that \textbf{the model trained on the full dataset does not achieve the highest performance}. Specifically, the model achieves the best performance when using only 70\% of the Big-Vul dataset, 80\% of the Devign dataset, and 40\% of the DiverseVul dataset, respectively. It indicates that some data instances may not be beneficial for boosting model performance. Instead, they may even harm the performance.

Our preliminary results confirm that \textbf{a non-trivial portion of natural training data can have a negative impact on model performance}. This finding highlights a critical challenge: improving performance requires not just more data, but a more discerning selection strategy. Active learning, a technique designed to reduce the significant data labeling burden, is a natural framework for this task. However, a conventional active learning strategy, often driven by uncertainty or diversity sampling, does not distinguish between informative samples and those hard-to-learn instances that can hinder the training process. This motivates our core research question:

\begin{tcolorbox}
\textit{How can we design a more effective selection strategy to guide active learning, enabling it to select samples based not only on their informativeness but also on their learnability to ensure a more efficient and stable training process?}
\end{tcolorbox}

\section{Methodology}\label{sec:method}

\subsection{Overview}
\begin{figure}[h]
    \centering
    \resizebox{.8\columnwidth}{!}{%
    \includegraphics{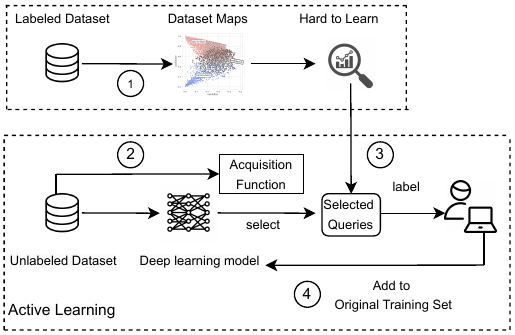}} 
    \caption{\textbf{Overview of Proposed Framework.}
    {\footnotesize 1) Identify hard-to-learn by constructing dataset maps from the labeled dataset. 2) Apply active learning to the unlabeled dataset, selecting informative samples. 3) Filter out selected samples by removing those similar to identified hard-to-learn before labeling. 4) Update training data by adding newly labeled samples to the original training set and retraining the model. Repeat steps 2, 3, and 4 iteratively.}}
    \label{fig:archi}
\end{figure}

To develop a more effective data selection strategy, we propose an approach that integrates dataset maps (described in Section~\ref{sec:dm}) into the standard active learning framework (described in Section~\ref{sec:al}) for vulnerability detection.

Figure~\ref{fig:archi} presents the overview of our proposed framework. The first step begins with constructing dataset maps from the labeled vulnerability dataset. By analyzing the model's training dynamics across multiple epochs, we iteratively identify the hard-to-learn samples that are consistently difficult to learn. To validate their impact, we compare the model's performance before and after removing these identified samples. If excluding the samples leads to improved training outcomes, the difficulty of these samples is used as an additional signal to guide the selection of new samples from the unlabeled pool. This systematic approach ensures that the identified samples are not merely hard to learn but genuinely detrimental to the model's learning process.

Secondly, once dataset maps are established and hard-to-learn examples are identified from the labeled dataset, we apply active learning to the unlabeled dataset drawn from the same underlying distribution, leveraging the identified hard-to-learn data to refine sample selection and mitigate their impact in the unlabeled data.
This selection follows a chosen acquisition function, such as uncertainty sampling or margin sampling, to identify the most informative instances for improving model performance. Thirdly, instead of directly using all selected samples, we compare them with the dataset maps’ identified hard-to-learn data. If an instance selected by active learning exhibits a high similarity to the hard-to-learn data identified in the dataset maps, it is considered likely to introduce noise or confusion into the learning process. To prevent the model from repeatedly training on misleading data, we systematically remove a portion of these overlapping instances before proceeding to the next iteration. 

Finally, this refined dataset is then used in the next active learning cycle, ensuring that each iteration selects samples that contribute positively to learning rather than reinforcing uncertainty.

By integrating our dataset maps-based method with active learning, we aim to enhance the selection process by eliminating hard-to-learn data. The benefits of our framework not only include improving model performance in vulnerability detection but also reducing the waste of expensive annotation efforts by preventing redundant or misleading instances from being repeatedly chosen.

\subsection{Dataset Maps}\label{sec:dm}
In the context of machine learning, the increased emphasis on data quantity has made it challenging to assess the quality of the data. Moreover, individual training examples differ significantly in their impact on model learning. Dataset maps provide a structured approach to analyzing dataset characteristics by leveraging training dynamics—statistical measures that track how a model learns individual examples over training~\cite{swayamdipta2020dataset}.

The methodology of dataset maps systematically categorizes data points based on two key properties: \textit{confidence} and \textit{variability}, which are computed across multiple training epochs.

The \textit{confidence} of a sample quantifies how consistently the model assigns a high probability to the correct label throughout training. Formally, for a data instance \( x_i \) with ground-truth label \( y^*_i \), confidence is defined as:

\begin{equation}
\hat{\mu}_i = \frac{1}{E} \sum_{e=1}^{E} p_{\theta^{(e)}}(y^*_i \mid x_i)
\label{equ:1}
\end{equation}

where \( p_{\theta^{(e)}}(y^*_i \mid x_i) \) represents the probability assigned to the correct label at epoch \( e \), and \( E \) is the total number of training epochs. High-confidence samples are those that the model consistently classifies correctly, whereas low-confidence samples remain uncertain or misclassified.  

The \textit{variability} of a sample measures fluctuations in confidence across training epochs. It is defined as the standard deviation:

\begin{equation}
\hat{\sigma}_i = \sqrt{\frac{\sum_{e=1}^{E} \left(p_{\theta^{(e)}}(y^*_i \mid x_i) - \hat{\mu}_i\right)^2}{E}}
\end{equation}

A high-variability sample is one for which the model oscillates between correct and incorrect predictions across epochs, indicating learnability challenges.  

By considering these two dimensions, dataset maps partition training data into three distinct regions:  

\begin{itemize}[wide=0pt]
    \item Easy-to-learn samples (high confidence, low variability): These examples are consistently learned and classified correctly. They contribute to model optimization but may not significantly enhance generalization.
    \item Ambiguous samples (high variability): The model frequently changes its prediction across training epochs for these samples, suggesting that they contain useful but challenging patterns that could improve out-of-distribution generalization.
    \item Hard-to-learn samples (low confidence, low variability): These examples are persistently misclassified, often due to noise, mislabeling, or inherent complexity.
\end{itemize}

Our preliminary experiment (described in Section~\ref{sec:motivation}) reveals the existence of samples that negatively impact training and reduce the final model's accuracy in vulnerability detection. To act on this insight, we leverage dataset maps, which provide a principled method for diagnosing and refining datasets by identifying samples that are particularly challenging for the model to learn. By systematically analyzing these hard-to-learn samples, we hypothesize that, by strategically managing these identified hard-to-learn samples within an active learning framework, we can significantly enhance training efficiency.

\subsection{Active Learning}\label{sec:al}
Active learning is a machine learning paradigm that aims to improve model performance while reducing labeling costs by strategically selecting the most informative data points for annotation. Instead of passively learning from randomly sampled data, an active learner iteratively queries an oracle (e.g., a human annotator) to label instances that the model finds most uncertain or impactful for refining its decision boundary~\cite{settles2009active}. This approach is particularly effective in scenarios where unlabeled data is abundant but annotation is expensive or time-consuming, as it prioritizes data points that contribute the most to model learning efficiency.
A critical component of active learning is the acquisition function, which determines the selection strategy for labeling. Existing approaches can be broadly categorized into uncertainty-based~\cite{beluch2018power,joshi2009multi,tong2001support} and diversity-based methods~\cite{bilgic2009link,gal2017deep,guo2010active}. Uncertainty-based selection prioritizes instances where the model's predictions are least confident, typically measured using entropy, margin sampling, or Gini impurity. The underlying intuition is that uncertain samples lie near the decision boundary, and incorporating them into training helps refine model decision-making. Diversity-based selection, such as K-means clustering, instead focuses on ensuring that selected instances span the entire feature space, capturing varied data distributions to improve generalization. By selecting cluster centroids as representatives, this method reduces redundancy in training data while ensuring that different data regions contribute to learning.

As a result, active learning has been considered a natural solution for vulnerability detection~\cite{moskovitch2008malicious,lu2014defect,xu2018cross,yu2019improving,hu2021towards,hu2024active} where security experts must manually review and label code for potential weaknesses, a process that is both time-consuming and requires specialized knowledge. By strategically selecting which code snippets to annotate, active learning reduces the labeling burden while improving detection accuracy. 
This assumption is challenged by findings that a model's performance can peak on a subset of the data, which suggests that not all examples are equally beneficial for learning. If an acquisition function focuses solely on a single metric like uncertainty, it may repeatedly select samples that do not efficiently guide the model toward a better decision boundary, leading to a less optimal learning path and plateaued performance. Addressing this limitation requires a more sophisticated approach that enhances the selection process with a deeper understanding of sample characteristics, ensuring that active learning truly accelerates progress toward a robust and effective model.

For this study, we adopt two acquisition functions from a recent work~\cite{hu2024active}, which have been demonstrated as the best performing ones in vulnerability detection. The DeepGini~\cite{feng2020deepgini} selects data with minimum Gini impurity, while K-means~\cite{sener2018active} is a clustering method to divide data into K groups and select the center of each group.

\section{Experimental Setting}
\subsection{Research Questions}


\begin{itemize}[wide=0pt]
    \item\textbf{RQ1:How effective is dataset maps in identifying hard to learn data in vulnerability detection?} 

    \textbf{Motivation}:
    Our preliminary experiments (described in Section~\ref{sec:motivation}) have shown that training a vulnerability detection model on the entire dataset does not always yield better performance compared to using a subset of the data. Therefore, we apply dataset maps to identify samples that are consistently hard to learn and potentially detrimental to model training. In this RQ, we aim to evaluate the effectiveness of dataset maps by comparing model performance when trained on the full dataset, a randomly selected subset, and the subset obtained by removing identified hard-to-learn.

    \item \textbf{RQ2: Can the identified hard-to-learn data be leveraged to boost the acquisition functions in active learning framework for vulnerability detection?}
    
    \textbf{Motivation}: Due to the lack of high-quality label nature of the vulnerability detection problem, active learning methods are often considered a solution in previous works~\cite{hu2021towards,li2012sample,moskovitch2008malicious,xu2018cross,yu2019improving}.
    In this RQ, our aim is to investigate whether the hard-to-learn of our approach can be used to improve existing active learning methods. More specifically, we evaluate the effectiveness of our proposed mechanism (described in Section~\ref{sec:method}) by comparing the performance of the standard active learning process with and without the integration of our mechanism.

    \item \textbf{RQ3: Can our proposed method perform consistently effective on a wide range of models and datasets?}
    
    \textbf{Motivation}: We aim to develop a method that can generally work well regardless of models and datasets. Hence, it is important to evaluate our method under a broader range of settings with different models and datasets.

\end{itemize}

\subsection{Datasets}\label{sec:dataset}

\begin{table}
\centering\caption{Statistics of the Datasets}
    \begin{tabular}{lccc}
        \toprule
        Dataset & \# Functions & \# Vulnerable & \#Non-Vulnerable \\
        \midrule
        BigVul  & 188.6K & 10.9K & 177.7K \\
        Devign   & 27.2K  & 12.4K & 14.8K  \\
        DiverseVul  & 330.4K  & 18.9K  & 311.5K  \\
        \bottomrule
    \end{tabular}   
    \label{tab:dataset_stats}
\end{table}

In this work, we consider the following datasets with different scales. They have been widely used in many prior works \cite{ding2024vulnerability,du2024generalization,hu2024active,liu2024pre,steenhoek2024dataflow,wu2022vulcnn}. Table~\ref{tab:dataset_stats} presents the statistics of the 3 datasets.

\begin{itemize}[wide=0pt]
    \item \textbf{Big-Vul}~\cite{fan2020ac} is collected from 348 Github projects. It consists of vulnerabilities collected from open-source projects across multiple programming languages, primarily C and C++, extracted from the National Vulnerability Database (NVD) and Common Vulnerabilities and Exposures (CVE) records.

    \item \textbf{Devign}~\cite{zhou2019devign} contains 27.2K functions collected from FFmpeg and QEMU, which are widely used open-source software projects—FFmpeg for multimedia processing and QEMU for hardware virtualization and emulation.

    \item \textbf{DiverseVul}~\cite{chen2023diversevul} is a dataset designed for deep learning-based vulnerability detection in C/C++ source code. It includes 18,945 vulnerable functions covering 150 distinct Common Weakness Enumerations (CWEs). These functions were extracted from 7,514 commits across 797 projects. 
\end{itemize}

\subsection{Models}

We consider the following three models which have been used in the existing vulnerability detection works~\cite{chen2023diversevul,hu2024active,liu2024pre,zhao2024coding,zhou2024large}.

\begin{itemize}[wide=0pt]
    \item \textbf{CodeBERT} is a bimodal model based on the BERT architecture, pre-trained on six code datasets covering multiple programming languages, including C,C++ and Python~\cite{feng2020codebert}. It tokenizes code into sequences, enabling it to capture semantic information similar to natural language processing models.
    \item \textbf{GraphCodeBERT} is a pre-trained code model that incorporates both sequences of tokens and data-flow information to learn code knowledge~\cite{guographcodebert}. This allows pre-trained code models to understand the structural information of programs and generate more precise code representations.
    \item \textbf{CodeT5} is a variant of the T5 model designed for code understanding tasks~\cite{wang2021codet5}. It follows the T5 architecture~\cite{raffel2020exploring} but integrates code-specific knowledge through pretraining on a large code corpus with natural language comments. CodeT5 employs Masked Span Prediction (MSP) and two auxiliary tasks, Identifier Tagging (IT) and Masked Identifier Prediction (MIP), to incorporate structural code information.

\end{itemize}

\subsection{Evaluation Metrics} 

The nature of the vulnerability detection problem is that the data is highly imbalanced, i.e., non-vulnerable code is much more than vulnerable code. To evaluate the model performance, we reuse the same metric as prior works~\cite{ding2024vulnerability,hu2024active,liu2024pre} as below:

\noindent$\bullet$ \textbf{F1-score (macro)}: The macro F1-score is often used in binary classification problems to measure the balance between precision and recall across all classes. It is computed by first calculating the F1-score for each class individually and then averaging them equally, regardless of class distribution. Given precision \( P_i \) and recall \( R_i \) for each class \( i \), the F1-score for that class is:
    
\[F1_i = \frac{2 P_i R_i}{P_i + R_i}\]
    
The \textbf{macro F1-score} is then obtained as:
\[\text{Macro-F1} = \frac{1}{C} \sum_{i=1}^{C} F1_i\]
where \( C \) is the total number of classes. Unlike the weighted F1-score, macro F1 treats all classes equally, making it useful for assessing model performance in imbalanced datasets.

\noindent$\bullet$ \textbf{Accuracy}: This metric measures the overall correctness of the model across all classes. It is defined as the ratio of correctly predicted observations (both true positives and true negatives) to the total number of observations. Our dataset consists of paired samples (vulnerable and benign function pairs), making the classes inherently balanced. Therefore, accuracy is particularly useful for evaluating the performance of models tested on a balanced dataset.

\section{Results}
\subsection{RQ1: Effectiveness in Identifying Hard to Learn}
\noindent\textbf{Experimental setting.}
To answer RQ1 and RQ2, we primarily focus on conducting experiments on the Big-Vul dataset (introduced in Section~\ref{sec:dataset}).
We randomly divide the dataset into two subsets with equal size, denoted as \textit{S1} and \textit{S2}.
Then, for each subset, we follow a standard ratio of 8:1:1 to split the subset into train, validation, and test sets. To address class imbalance in the training data, we include all available positive samples and randomly select an equal number of negative samples. This dataset setting follows the setup used in Yu's paper~\cite{yu2019improving} for evaluating active learning. 
Meanwhile, the validation and test sets retain the natural distribution of the dataset to reflect real-world conditions, where non-vulnerable code is significantly more prevalent than vulnerable code.
We process \textit{S1} and \textit{S2} in the same manner. Then, we use \textit{S1} for answering RQ1 and use the identified hard-to-learn data to perform active learning on \textit{S2} for answering RQ2.

We train the CodeBERT model on \textit{S1} for 10 epochs. During training, we record the model's predicted logits for each sample at every epoch. Since the first two epochs alone often do not provide stable and sufficient information to assess how a sample's logits evolve over time, we begin tracking two key statistics from the 3rd to the 10th epoch: the number of times a sample has been correctly classified up to that point and instances where a previously correct prediction is forgotten. Using these dynamics, we compute a difficulty metric based on confidence and variability, where confidence measures the certainty in model's predictions, and variability captures the frequency of prediction changes across epochs (detailed in Section~\ref{sec:dm}). 
After calculating confidence and variability for all samples, we define hard-to-learn as those whose confidence and variability values fall below the threshold values of 0.3 and 0.4, respectively. The threshold values of 0.3 for confidence and 0.4 for variability are selected based on empirical observations of dataset maps across multiple experimental runs. These parameters are determined by enumerating configurations and assessing their impact on the validation data. To maintain consistency and control variables in subsequent experiments, we use the same threshold settings across all the trials. Hard-to-learn data are identified separately at the 3rd and 10th epochs, resulting in different sets at different training stages.

To determine which set of hard-to-learn data has the most negative impact on training, we
systematically remove them from the full \textit{S1} dataset and retrain the model on the remaining data. To ensure that the observed improvements are driven by the dataset maps' selection rather than the removal process itself, we conduct an additional experiment where, at each epoch, we randomly remove the same number of samples five times and retrain the model under the same parameter settings.

\begin{figure}[]
    \centering
    \includegraphics[width=.6\linewidth]{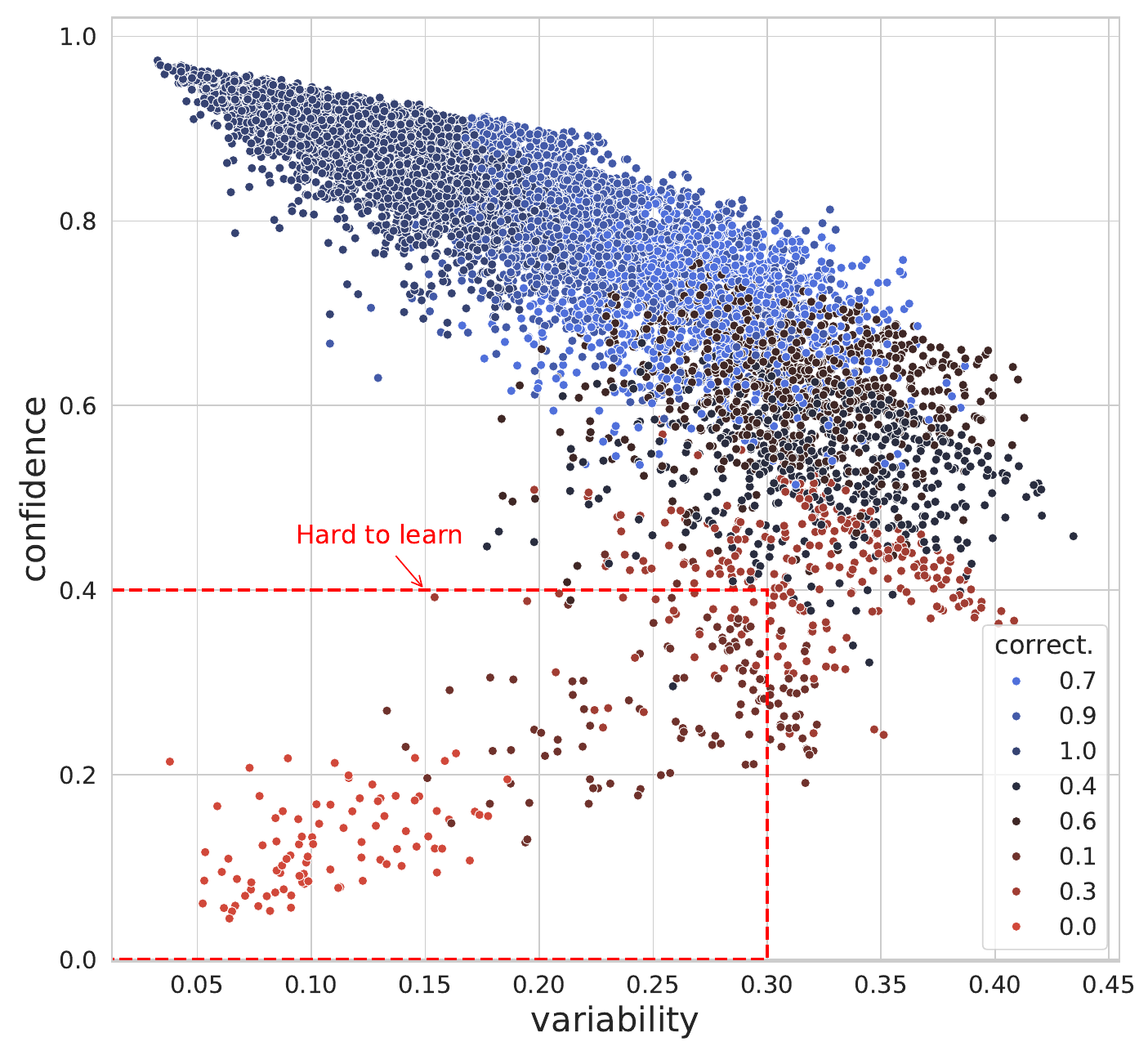}
    \caption{An Example of Dataset Maps Visualization. Blue dots represent consistently correct samples (correctness $\geq 0.9$), black/brown indicate moderate correctness ($0.4 \text{–} 0.7$), and red denotes frequently misclassified samples (correctness $\leq 0.3$).}
    \label{fig:rq1_map}
    \vspace{-10pt}
\end{figure}

\noindent\textbf{Results}: 
Figure \ref{fig:rq1_map} presents an example of training dynamics visualization computed from all data points at epoch 7, illustrating how CodeBERT's confidence and variability evolved during training. 

Based on predefined threshold values, data points within the highlighted red box represent hard-to-learn data indicating samples that exhibit frequent misclassifications or unstable learning behavior. The correctness in figure is an outcome-based measure that directly reflects how often the model classifies a sample correctly. Similar visualizations are generated for each epoch from epoch 3 to epoch 10, allowing a comprehensive analysis of how these samples impact model learning across different training stages.

\begin{figure}[]
    \centering
    \resizebox{.8\columnwidth}{!}{%
    \includegraphics{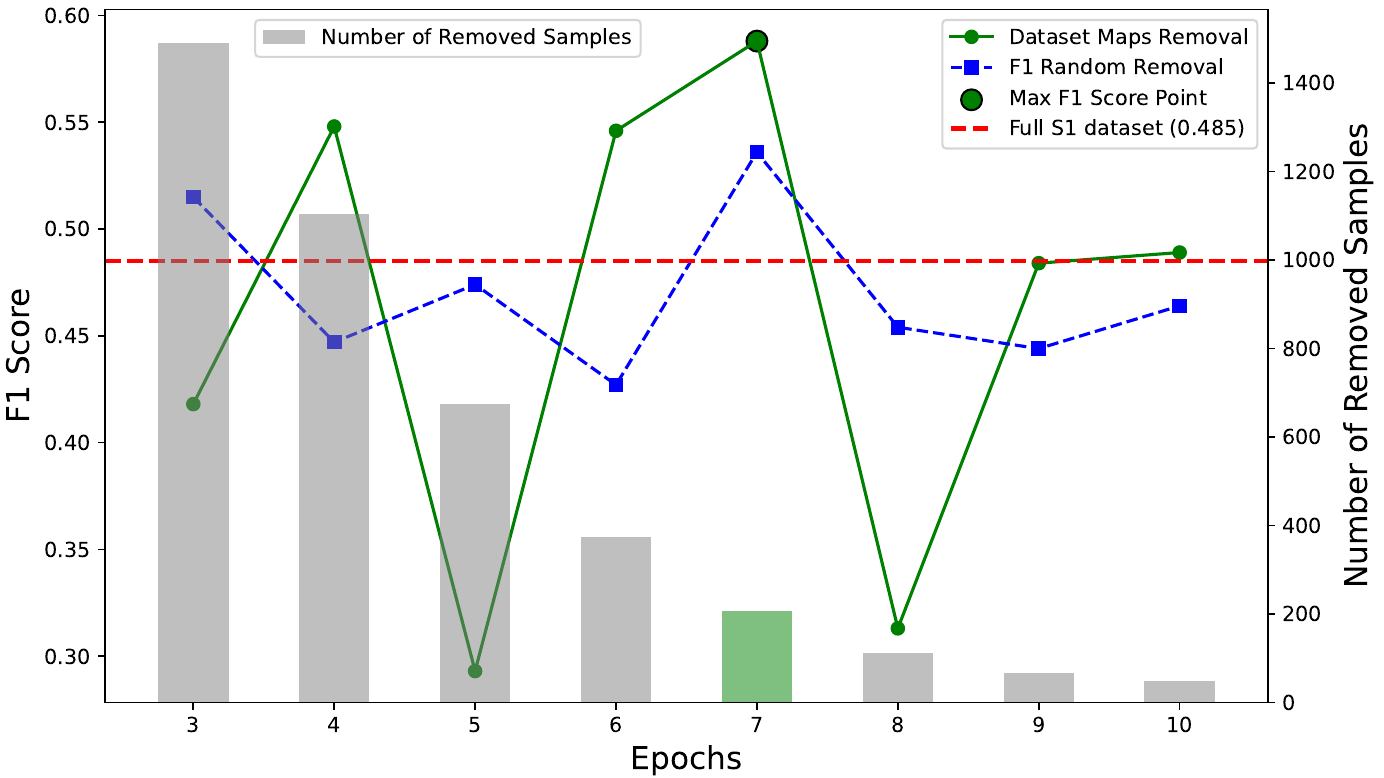}}
    \caption{Hard-to-Learn Identified from Epochs 3 to 10 and Their Impact on Model Performance vs. Random Removal.}
    \label{fig:rq1_f1}
    \vspace{-10pt}
\end{figure}

Figure~\ref{fig:rq1_f1} illustrates the number of hard-to-learn data identified across epochs 3 to 10 and the corresponding model performance after their removal. It also presents a comparison with the model trained on data where the same number of samples were randomly removed. The detailed numerical results for all experiments are saved in the replication package~\cite{badseeds}.

From the results, we observe that removing the 207 hard-to-learn samples identified at epoch 7 yields the highest performance gain, with an F1 score of 0.588. This represents a 21.23\% improvement over the baseline model trained on the full \textit{S1} dataset (F1 = 0.485). In contrast, randomly removing 207 samples results in an F1 score of 0.536, which is better than the full dataset but still inferior to the dataset map-based removal. This highlights that the hard-to-learn data identified at epoch 7 have the most severe negative impact on model training, reinforcing the effectiveness of dataset maps.

\begin{tcolorbox}
\textbf{Answer summary to RQ1}: Dataset maps effectively identify hard-to-learn data that negatively impact model performance. The results show that removing 207 hard-to-learn samples at epoch 7 achieves the highest performance gain, with an F1 score of 0.588, a 21.23\% increase compared to training on the full \textit{S1} set (F1 = 0.485).
\end{tcolorbox}

\subsection{RQ2: Effectiveness in Enhancing Active Learning}
\textbf{Experimental setting}:
As introduced in the experimental setting of RQ1, for answering RQ2, we conduct experiments to evaluate whether the hard-to-learn data identified from \textit{S1} in RQ1 can improve the acquisition function in active learning. The 207 hard-to-learn samples identified in RQ1 are utilized to assess their influence on active learning performance in RQ2. Specifically, we focus on two state-of-the-art active learning methods for vulnerability detection: DeepGini and K-means, identified in Hu et al. work \cite{hu2024active}. We use \textit{S2} to perform active learning experiments on CodeBERT with two settings:

\begin{itemize}[wide=0pt]
    \item \textbf{Standard active learning}. Following the same experimental setup described in Hu's paper~\cite{hu2024active}, we randomly sample 500 data points from \textit{S2} as the initial seed set to train a base model, ensuring that the initial model has minimal prior knowledge of the dataset. We set the labeling budget to 100 samples per iteration, approximately 1\% of the training set, and perform 10 iterations of active learning. At each iteration, the trained model evaluate the remaining \textit{S2} dataset, and an acquisition function (DeepGini or K-means) is used to select 100 additional data points to expand the seed set. The model is then retrained with the updated dataset.
    \item \textbf{Hard-to-learn informed active learning}. In this setting, we incorporate the signal of hard-to-learn samples identified in RQ1 into the acquisition function optimization. For a fair comparison, we use the same initial seed set of 500 data points as in the first setting, the model evaluates the remaining \textit{S2} dataset, and the acquisition function selects 125 candidate data points at each iteration. For these 125 samples, we computed the cosine similarity of each data point with the hard-to-learn data from RQ1. The top 20\% of samples (25 data points) with the highest similarity scores are discarded, and the remaining 100 data points are added to the seed set which is the same size with the standard active learning setting. This iterative process is repeated for 10 iterations.
\end{itemize}

The performance of the models from both settings is compared in each iteration, using F1-score to assess the effectiveness of active learning. The comparison aims to determine whether excluding data points similar to the hard-to-learn data leads to better model performance and optimizes the acquisition function. As a baseline, we also follow the approach in Hu et al.'s work~\cite{hu2024active}, in which 100 data points were randomly selected at each iteration instead of using an acquisition function. We also do the random selection 5 times to avoid bias.

\begin{figure}[]
    \centering
    \begin{minipage}{0.48\textwidth}
        \centering
        \includegraphics[width=\linewidth]{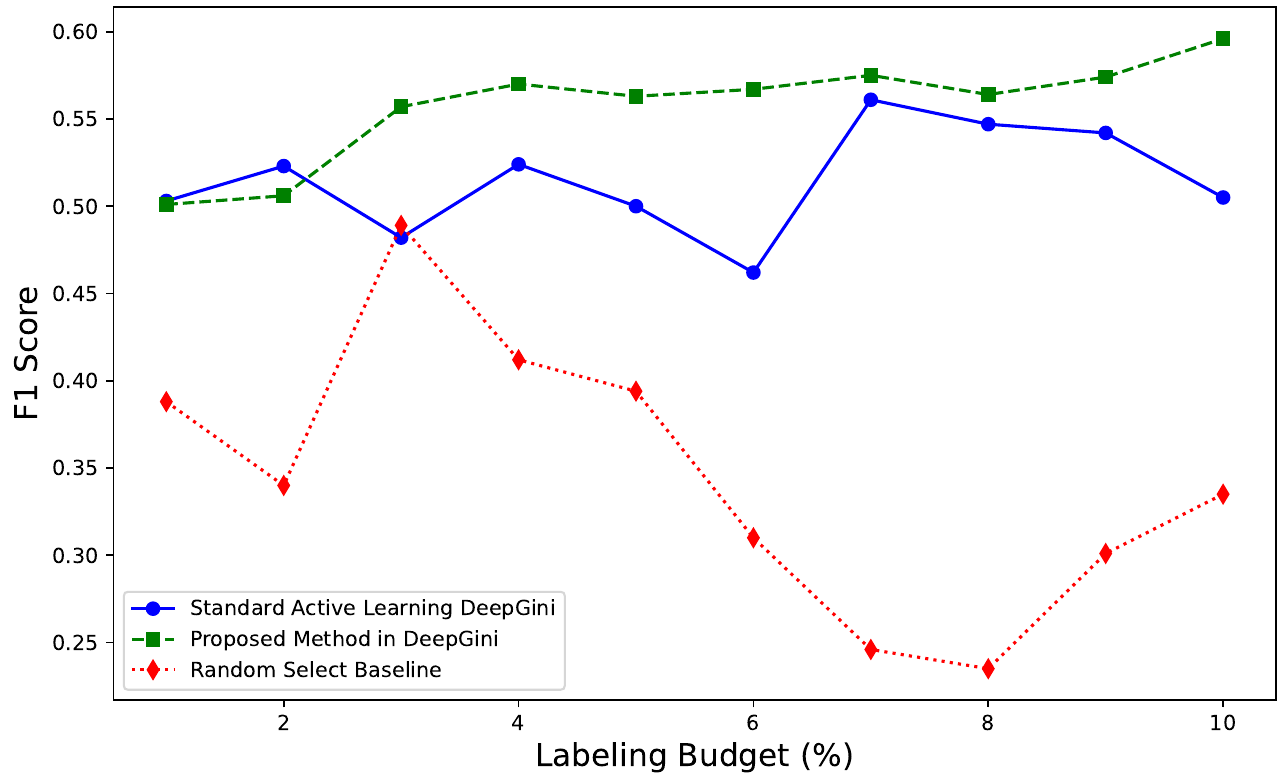}
        \caption{CodeBERT trained on Big-Vul (DeepGini): \\F1 Score Across Labeling Budgets}
        \label{fig:rq2_deep}
    \end{minipage}
    \hfill
    \begin{minipage}{0.48\textwidth}
        \centering
        \includegraphics[width=\linewidth]{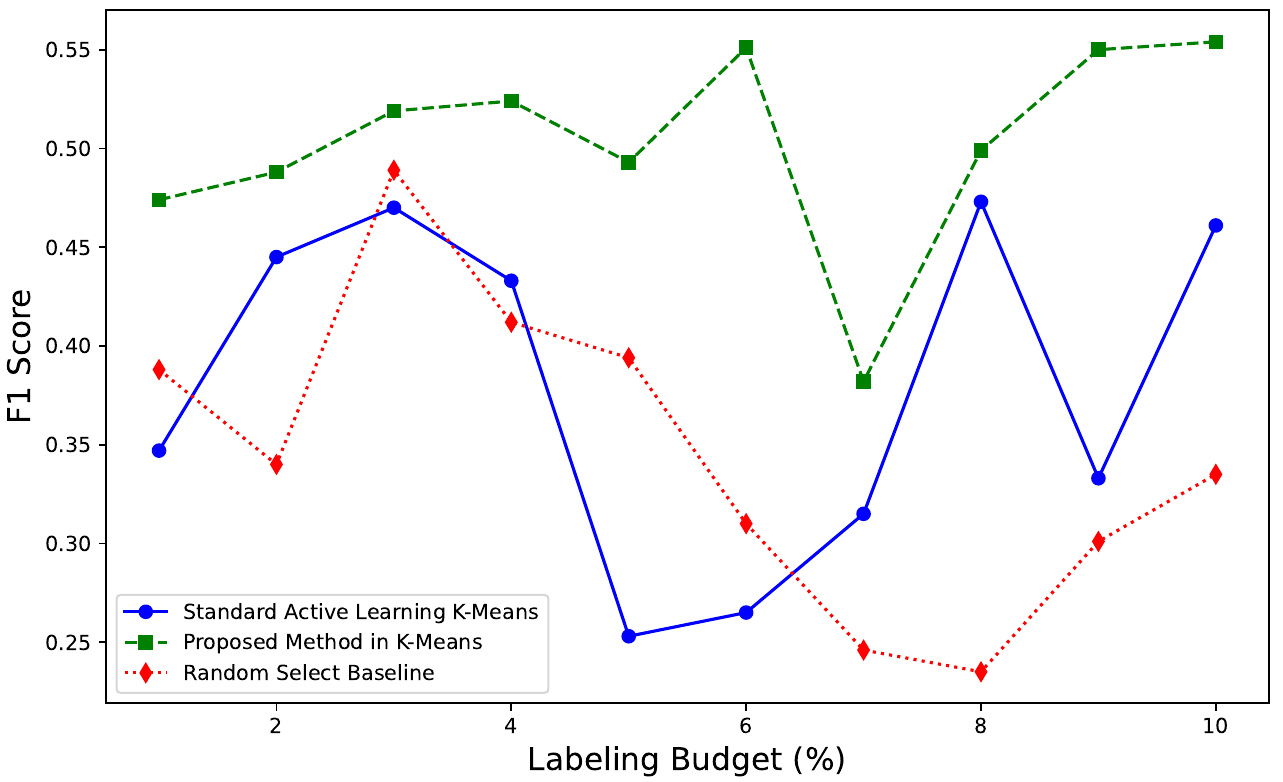}
        \caption{CodeBERT trained on Big-Vul (K-Means): \\F1 Score Across Labeling Budgets}
        \label{fig:rq2_km}
    \end{minipage}
\end{figure}

\noindent\textbf{Results}: 
Our results for DeepGini and K-Means are presented in Figure \ref{fig:rq2_deep} and Figure \ref{fig:rq2_km}, respectively. 
To evaluate the statistical significance of the observed performance differences, we follow Hu et al.'s approach~\cite{hu2024active} and conduct a paired t-test~\cite{owen1965power}, as shown in Table \ref{tab:rq2}. Based on the results, we observe that the proposed method generally outperforms both random selection and standard active learning across multiple iterations. The numerical F1 scores for all 10 iterations are recorded in the replication package~\cite{badseeds}. By comparing the average F1 scores over 10 iterations, the DeepGini-based proposed method improves by 61.54\% over random selection, while the K-Means-based proposed method achieves a 45.91\% improvement. Additionally, compared to standard active learning, the proposed method enhances performance by 8.23\% for DeepGini and 32.65\% for K-Means. 

\noindent\textbf{Enhancement on DeepGini}. 
The DeepGini-based proposed method significantly outperforms both random selection and standard DeepGini. Compared to random selection, it achieves a t-statistic of 7.4573 with \( p < 0.01 \), and against standard DeepGini, it achieves a t-statistic of 3.3019 with \( p < 0.01 \). These results indicate that while standard DeepGini offers a clear advantage over random selection, integrating our analysis of hard-to-learn samples always enhances the selection process (except 20\% labeling budget), leading to superior model performance.

\noindent\textbf{Enhancement on K-Means}.
For K-Means, the proposed method outperforms both random selection and standard K-Means, achieving a t-statistic of 6.2464 and 4.2793, respectively. The corresponding \( p \)-values (\( p < 0.01 \) for both comparisons) indicate that these improvements are statistically significant. These results highlight that incorporating dataset maps into active learning effectively optimizes the selection process, ensuring that more valuable samples contribute to training and enhancing overall model robustness.

Overall, from Figure \ref{fig:rq2_deep} and Figure \ref{fig:rq2_km}, the proposed DeepGini method achieves a peak F1-score of 0.596, while the proposed K-Means method reaches 0.554. A key observation is that DeepGini demonstrates a more stable improvement trend, particularly after the initial iterations, where it consistently outperforms the standard method. In contrast, K-Means exhibits more fluctuation across iterations, likely due to the nature of clustering-based selection, which can introduce variability in sample acquisition. It indicates that DeepGini benefits more from dataset maps in filtering out misleading samples and selecting informative ones, allowing for more stable learning progression. 

\begin{table}[]
    \centering
    \caption{T-Test Results: Proposed Method vs. Standard Active Learning and Random Selection (CodeBERT on Big-Vul)}
    \resizebox{.5\linewidth}{!}{
    \begin{tabular}{llcc}
        \toprule
        & & \textbf{DeepGini} & \textbf{K-means} \\
        \midrule
        \multirow{2}{*}{\shortstack{Proposed Method vs \\ Random Selection}} & t-statistic & 7.4573 &  6.2464\\
        & P-value & 0.0001 & 0.0002 \\
         \midrule
        \multirow{2}{*}{\shortstack{Proposed Method vs \\ Standard Active Learning}} & t-statistic & 3.3019 & 4.2793 \\
        & P-value & 0.0092 & 0.0021 \\
        \bottomrule
    \end{tabular}
    }
    \label{tab:rq2}
\end{table}

\begin{tcolorbox}
\textbf{Answer summary to RQ2}: Based on our proposed integration mechanism, the identified hard-to-learn samples can be leveraged to optimize the acquisition function in the active learning framework for vulnerability detection. By filtering out samples that exhibit unstable learning dynamics, the proposed method refines the selection process, ensuring that the model focuses on more valuable instances.
\end{tcolorbox}

\subsection{RQ3: Generalizability of Our Proposed Method}

\textbf{Experiment setting}:
To evaluate the applicability of our proposed approach across different models and datasets, we extend our experiments by incorporating GraphCodeBERT and CodeT5 as additional LLM-based vulnerability detectors and by testing on Devign and DiverseVul datasets alongside Big-Vul. 

For multi-model validation, we replace CodeBERT with GraphCodeBERT and CodeT5 while maintaining the same experimental setup in RQ1 and RQ2. Each model generates its own dataset maps to identify hard-to-learn samples, which are then used in active learning experiments. For multi-dataset validation, we use CodeBERT and apply the approach to Devign and DiverseVul while preserving class distributions in test and validation sets. This ensures a consistent evaluation framework, allowing us to assess the generalizability of our method across different models and datasets.

\begin{figure*}
    \centering 
    \begin{subfigure}{0.48\textwidth} 
        \includegraphics[width=\linewidth]{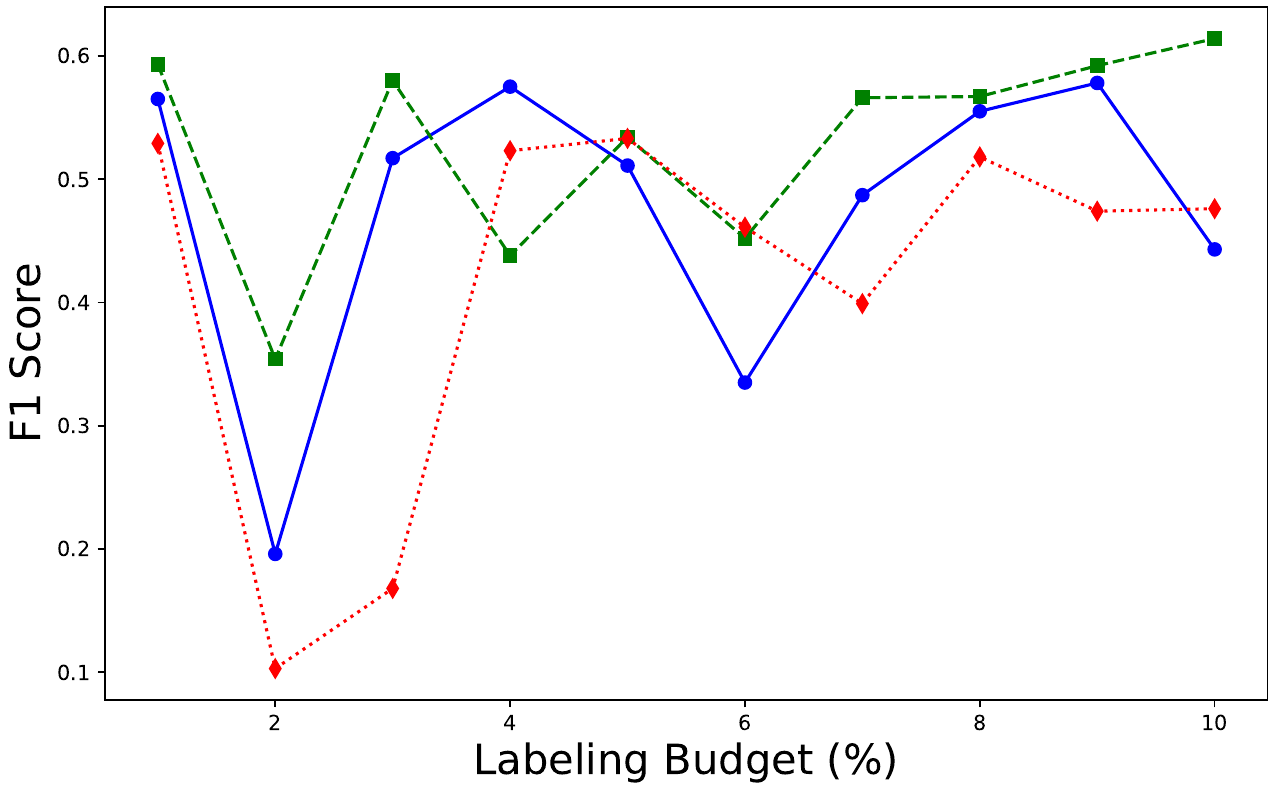}
        \captionsetup{justification=centering}
        \caption{GraphCodeBERT trained on Big-Vul (DeepGini)}
    \end{subfigure}
    \hfill
    \begin{subfigure}{0.48\textwidth}
        \includegraphics[width=\linewidth]{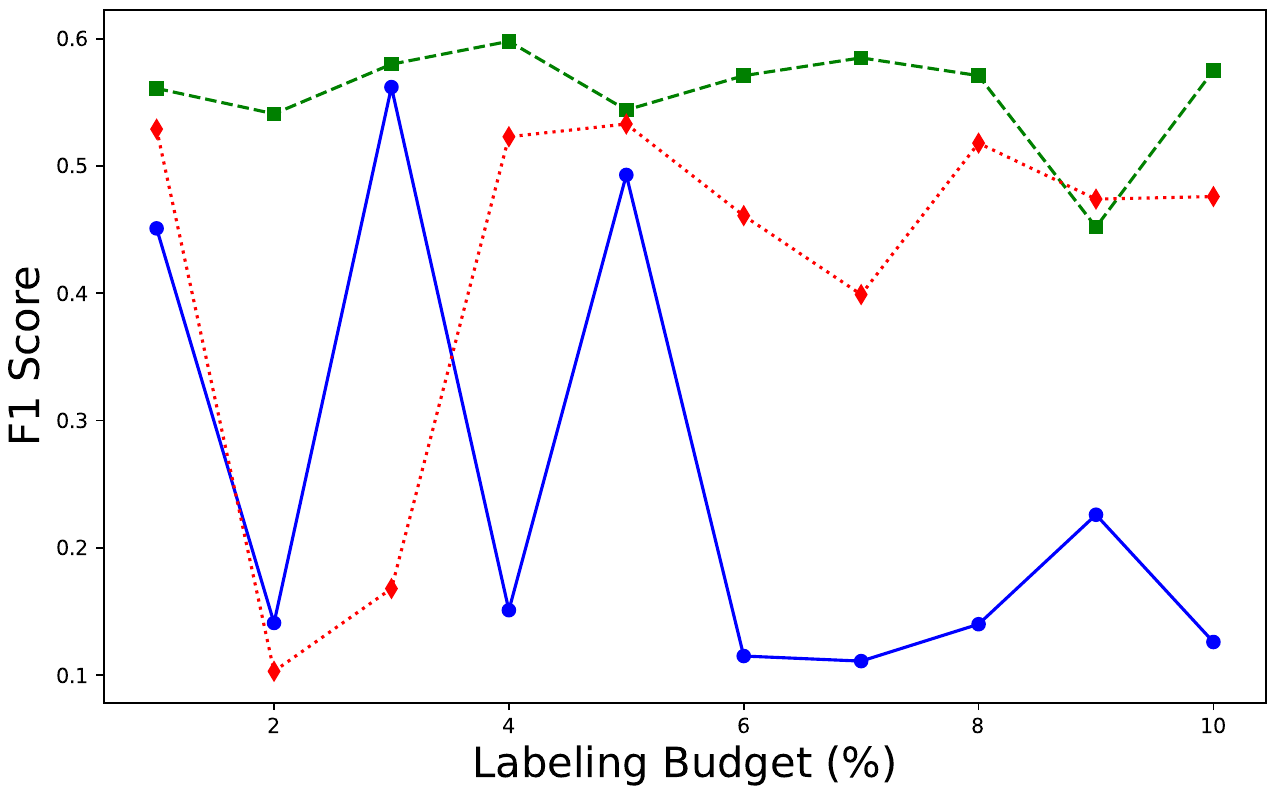}
        \captionsetup{justification=centering}
        \caption{GraphCodeBERT trained on Big-Vul (K-Means)}
    \end{subfigure}

    \begin{subfigure}{0.48\textwidth} 
        \includegraphics[width=\linewidth]{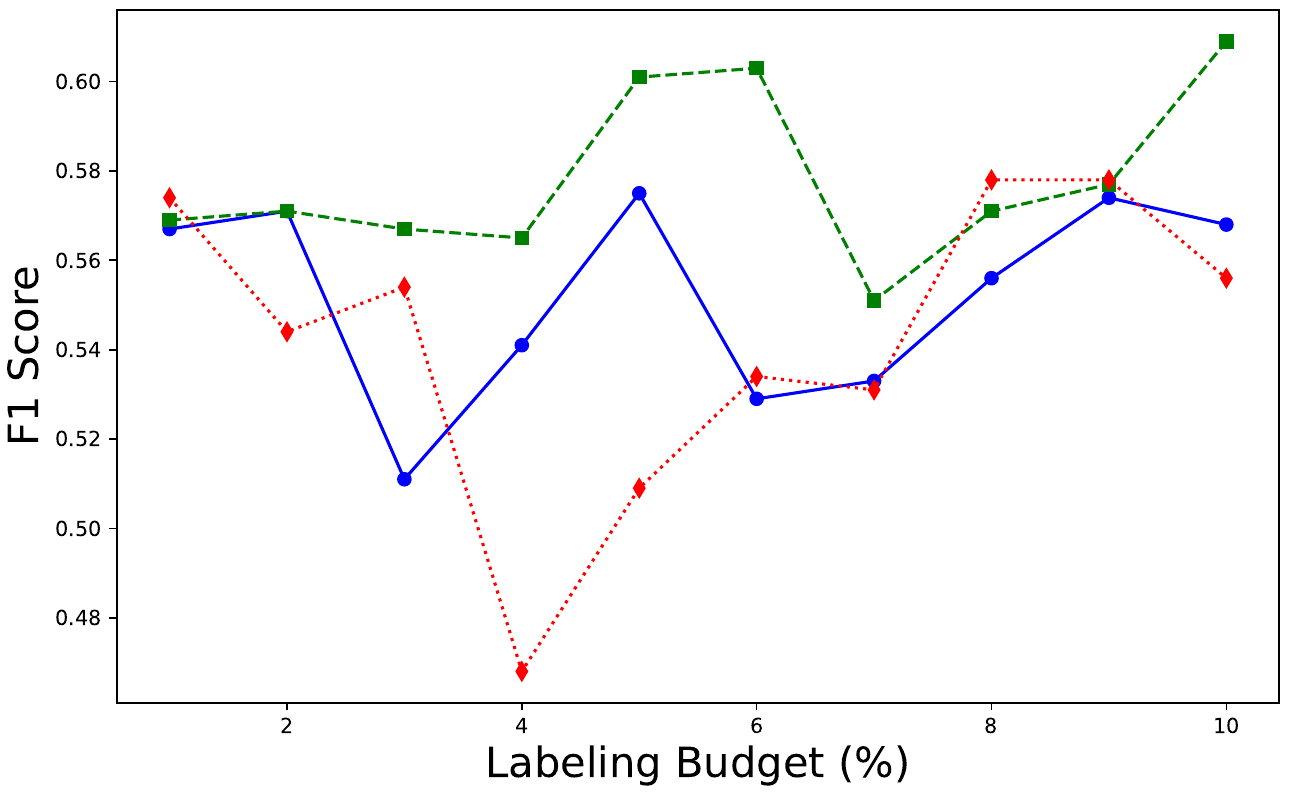}
        \captionsetup{justification=centering}
        \caption{CodeT5 trained on Big-Vul (DeepGini)}
    \end{subfigure}
    \hfill
    \begin{subfigure}{0.48\textwidth}
        \includegraphics[width=\linewidth]{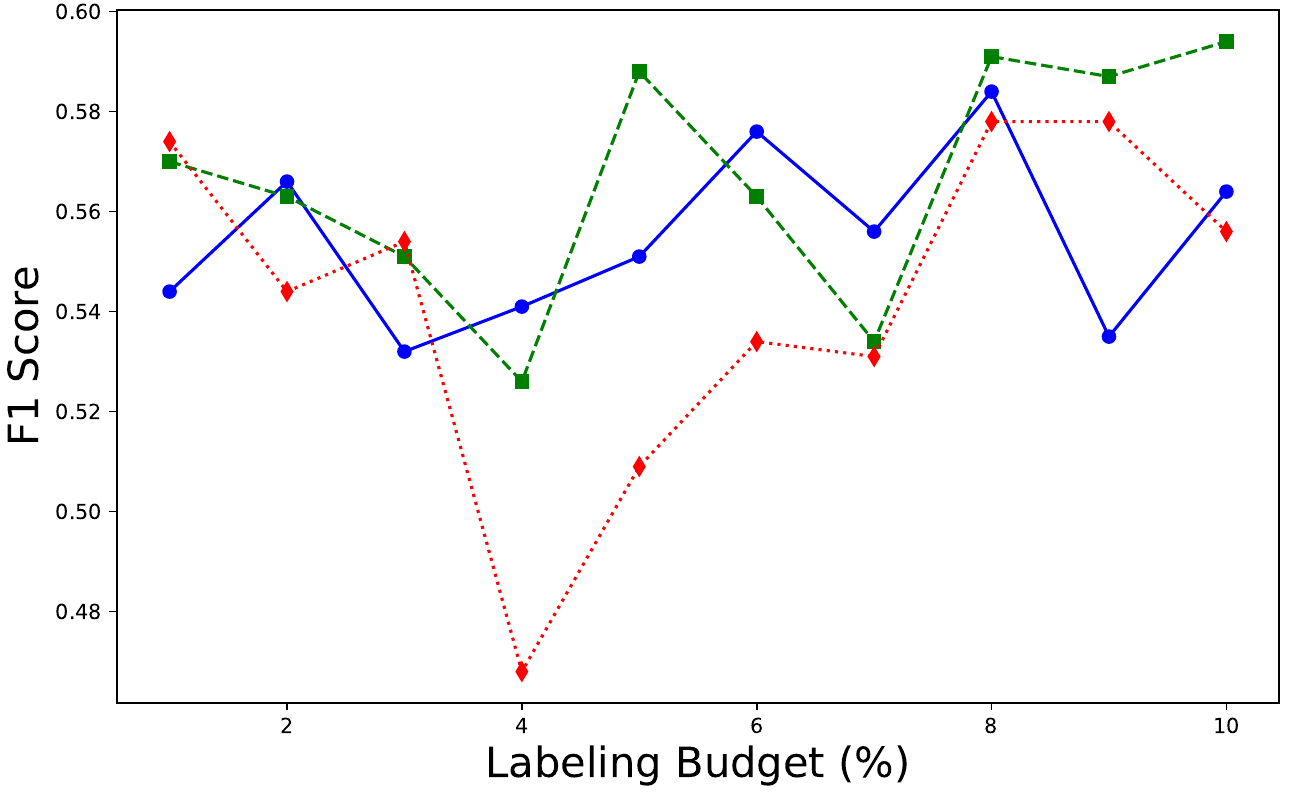}
        \captionsetup{justification=centering}
        \caption{CodeT5 trained on Big-Vul (K-Means)}
    \end{subfigure}

    \begin{subfigure}{0.48\textwidth}
        \includegraphics[width=\linewidth]{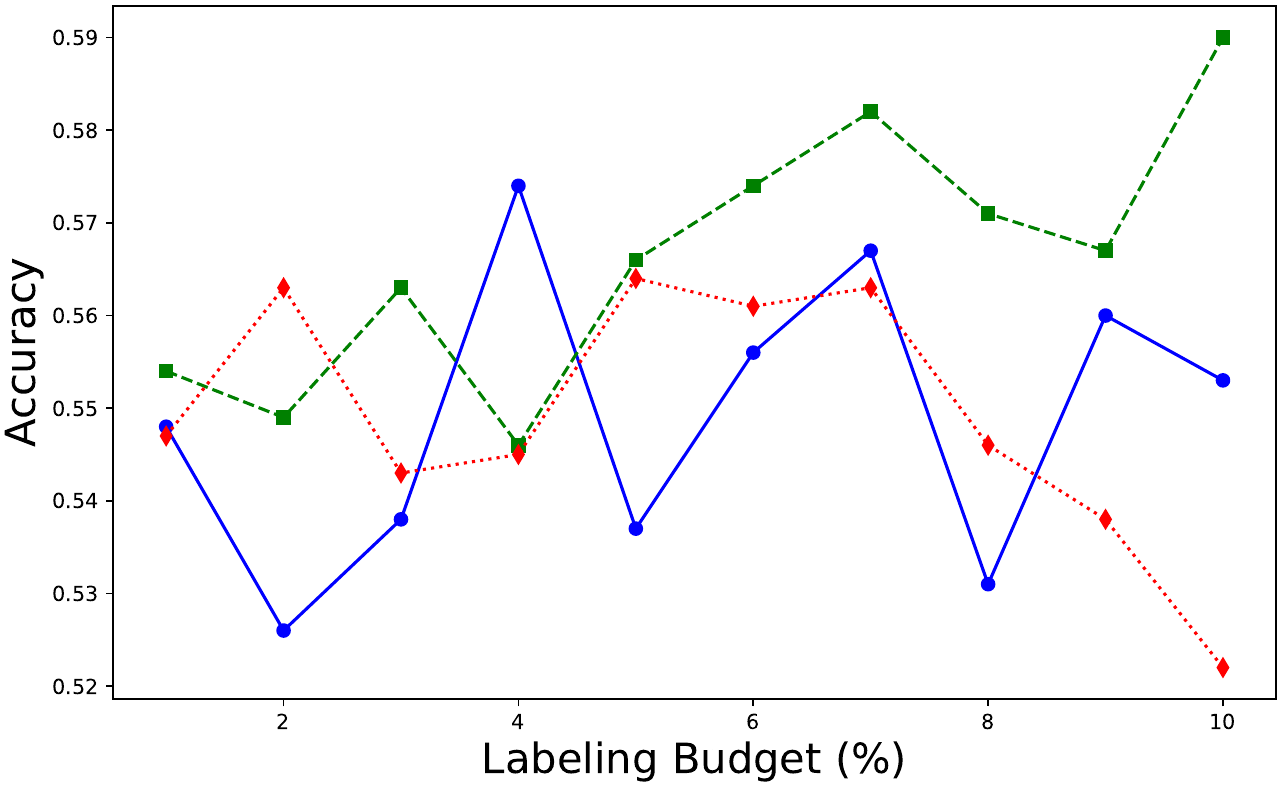}
        \captionsetup{justification=centering}
        \caption{CodeBERT trained on Devign (DeepGini)}
    \end{subfigure}
    \hfill
    \begin{subfigure}{0.48\textwidth}
        \includegraphics[width=\linewidth]{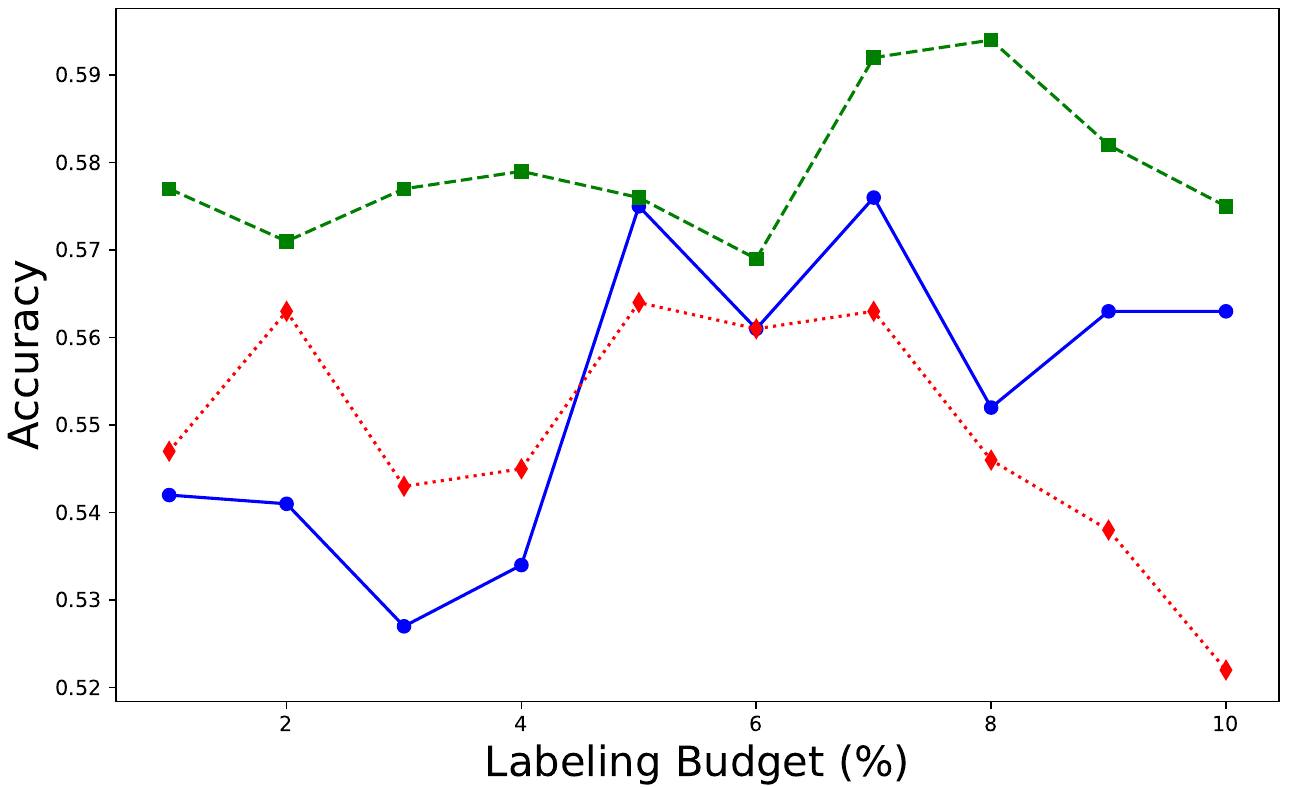}
        \captionsetup{justification=centering}
        \caption{CodeBERT trained on Devign (K-Means)}
    \end{subfigure}

    \begin{subfigure}{0.48\textwidth}
        \includegraphics[width=\linewidth]{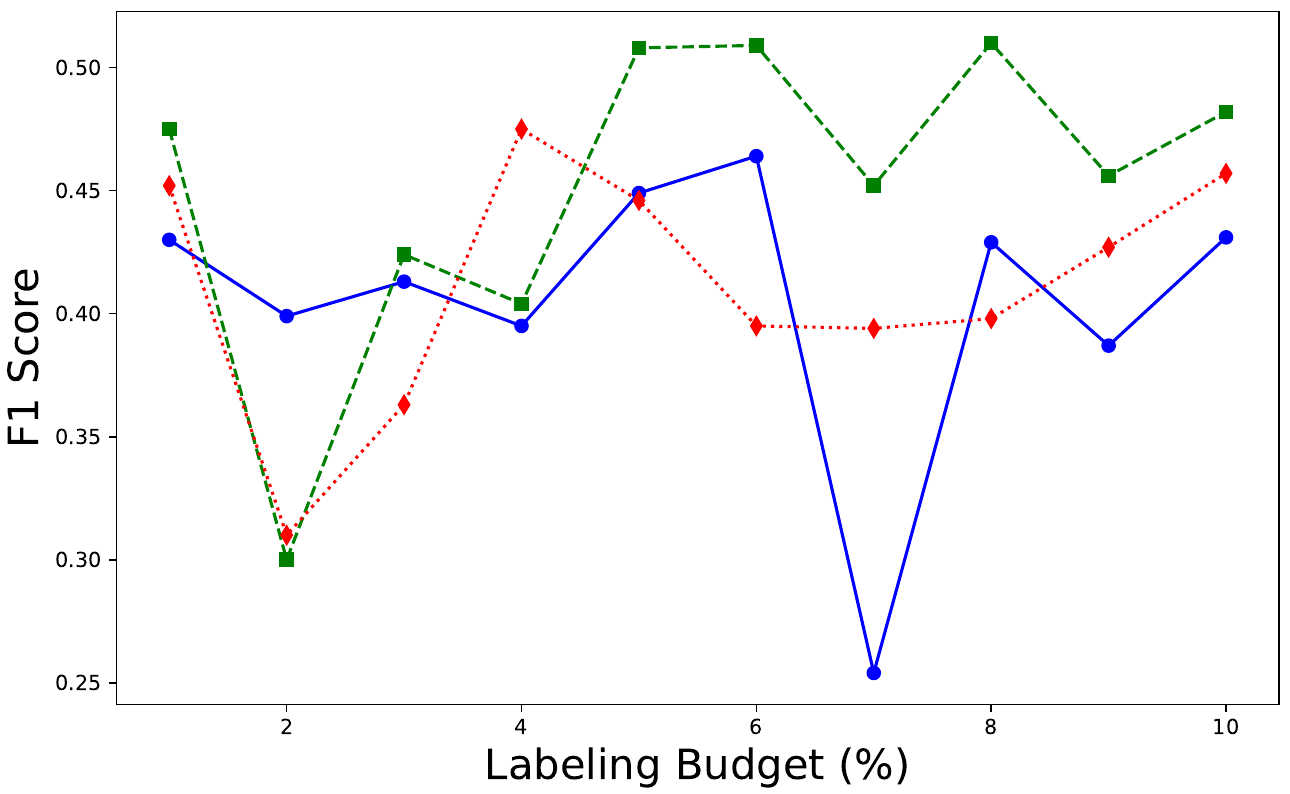}
        \captionsetup{justification=centering}
        \caption{CodeBERT trained on DiverseVul (DeepGini)}
    \end{subfigure}
    \hfill
    \begin{subfigure}{0.48\textwidth}
        \includegraphics[width=\linewidth]{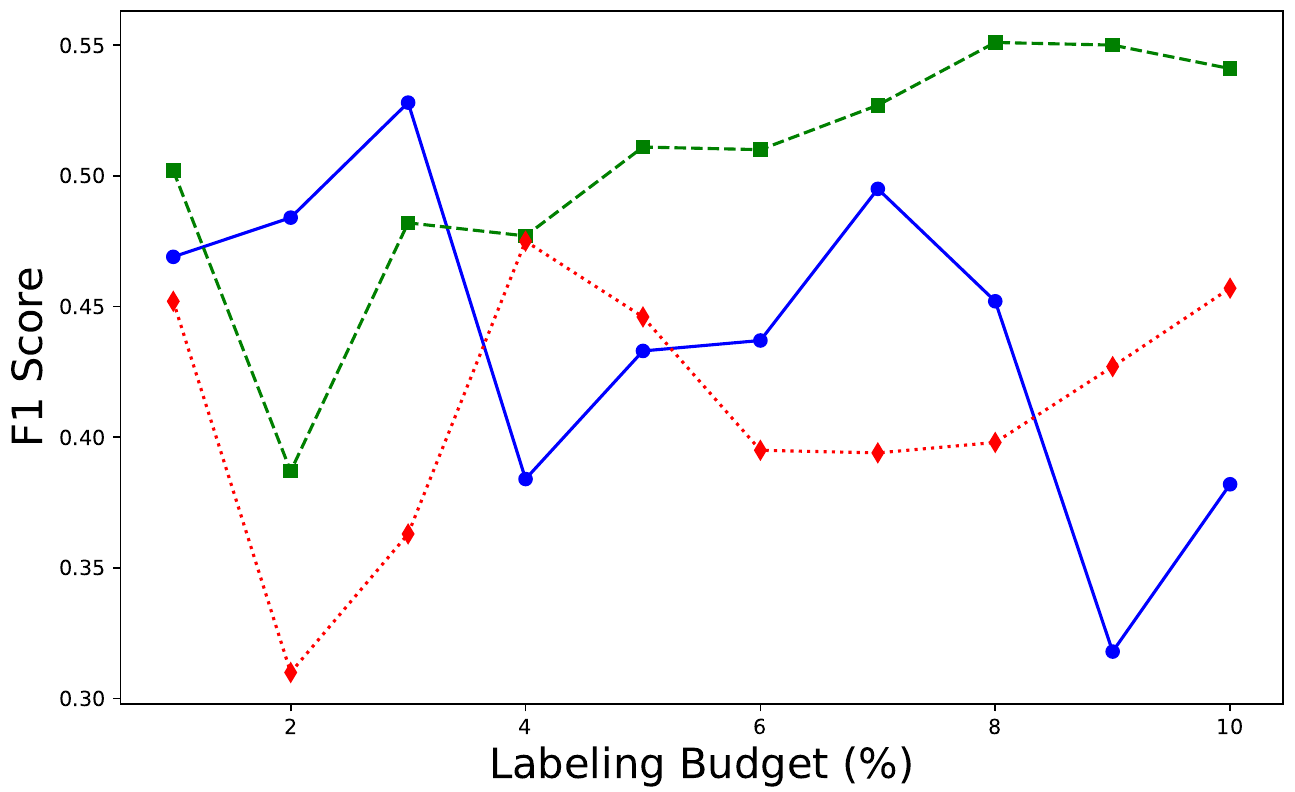}
        \captionsetup{justification=centering}
        \caption{CodeBERT trained on DiverseVul (K-Means)}
    \end{subfigure}

    \caption{\textbf{Performance Comparison of the Proposed Method Across Multiple Models and Datasets. }{\footnotesize The \textcolor{mydarkgreen}{green} line represents the proposed method applied to DeepGini or K-Means, the \textcolor{blue}{blue} line denotes standard active learning (DeepGini or K-Means), and the \textcolor{red}{red} line indicates the random selection baseline.}}
    \label{fig:rq3}
\end{figure*}

\noindent\textbf{Results}: 
Following the same procedure as in RQ1, we identify the number of hard-to-learn samples for each model and dataset: 197 for GraphCodeBERT on BigVul, 45 for CodeT5 on BigVul, 148 for CodeBERT on Devign, and 397 for CodeBERT on DiverseVul. These identified hard-to-learn samples are subsequently used in the next step to enhance active learning in the proposed method. All dataset maps results, along with detailed numerical values for the following experiments, are available in the replication package~\cite{badseeds}.

\begin{table}[H]
    \centering
    \caption{T-Test Results: Proposed Method vs. Random Selection (Multi-Model and Multi-Dataset)}
    \resizebox{.5\linewidth}{!}{
    \begin{tabular}{llcc}
        \toprule
        \textbf{Model (Dataset)} & & \textbf{DeepGini} & \textbf{K-means} \\
        \midrule
        \multirow{2}{*}{GraphCodeBERT (Big-Vul)} & t-statistic & 2.4389 & 2.7349\\
        & P-value & 0.0374 & 0.0230\\
        \midrule
        \multirow{2}{*}{CodeT5 (Big-Vul)} & t-statistic 
        & 2.8704 & 2.7876 \\
        & P-value & 0.0185 & 0.0211 \\
         \midrule
        \multirow{2}{*}{CodeBERT (Devign)} & t-statistic 
        & 2.4369 & 5.8412 \\
        & P-value & 0.0376 & 0.0002 \\
         \midrule
        \multirow{2}{*}{CodeBERT (DiverseVul)} & t-statistic &2.3112 & 6.4233 \\
        & P-value &0.0461 & 0.0001 \\
        \bottomrule
    \end{tabular}
    }
    \label{tab:rq3_tb1}
\end{table}

\begin{table}[h]
    \centering
    \caption{T-Test Results: Proposed Method vs. Standard Active Learning, (Multi-Model and Multi-Dataset)}
    \resizebox{.5\linewidth}{!}{
    \begin{tabular}{llcc}
        \toprule
        \textbf{Model (Dataset)} & & \textbf{DeepGini} & \textbf{K-means} \\
        \midrule
        \multirow{2}{*}{GraphCodeBERT (Big-Vul)} & t-statistic & 1.8845 & 5.2317 \\
        & P-value & 0.0921  & 0.0005 \\
        \midrule
        \multirow{2}{*}{CodeT5 (Big-Vul)} & t-statistic 
        & 3.3440 & 1.5022 \\
        & P-value & 0.0086& 0.1673\\
         \midrule
        \multirow{2}{*}{CodeBERT (Devign)} & t-statistic 
        & 2.7887 & 4.8154 \\
        & P-value & 0.0211 & 0.0009 \\
         \midrule
        \multirow{2}{*}{CodeBERT (DiverseVul)} & t-statistic & 2.0169 & 2.2062 \\
        & P-value & 0.0745& 0.0548\\
        \bottomrule
    \end{tabular}
    }
    \label{tab:rq3_tb2}
\end{table}

\noindent\textbf{ Generalizability on models} is presented in the first two rows of Figure \ref{fig:rq3}. We find that both GraphCodeBERT and CodeT5 show significant performance improvements when integrating hard-to-learn signals into active learning. As shown in Table \ref{tab:rq3_tb1}, the t-test results confirm statistically significant improvements over random selection across both DeepGini and K-Means. 
Furthermore, Table \ref{tab:rq3_tb2} shows that, compared to standard active learning, CodeT5 demonstrates a strong advantage with a t-statistic of 3.3440 in DeepGini, while GraphCodeBERT achieves a clear improvement under K-Means (t = 5.2317).
However, for GraphCodeBERT with DeepGini and CodeT5 with K-Means, the differences are less pronounced, suggesting that the effectiveness of dataset maps may vary depending on the model and acquisition function.

\noindent\textbf{ Generalizability on datasets} is presented
 in the last two rows of Figure \ref{fig:rq3}. The proposed method consistently outperforms random selection across all datasets, as reflected in Table \ref{tab:rq3_tb1}, where the improvement is particularly strong in DiverseVul, with K-Means achieving a t-statistic of 6.4233, highlighting the substantial impact of filtering hard-to-learn data. Compared to standard active learning, the results in Table \ref{tab:rq3_tb2} indicate significant gains on Devign, where both DeepGini and K-Means yield strong t-statistics (2.7887 and 4.8154, respectively). However, in DiverseVul, the improvement over standard active learning is less consistent, with higher p-values, suggesting that the model's ability to leverage dataset maps may depend on dataset characteristics.

Overall, our results indicate that the proposed method consistently outperforms random selection across all models and datasets, demonstrating its effectiveness in improving active learning.

\begin{tcolorbox}
\textbf{Answer summary to RQ3}: 
Our proposed method can consistently enhance active learning in vulnerability detection in a diverse range of settings in terms of acquisition functions, machine learning models, and datasets.
\end{tcolorbox}

\section{Discussion}

\subsection{Towards better-and-also-stable active learning in vulnerability detection}
In Hu et al.'s paper~\cite{hu2024active}, their original active learning study applied DeepGini and K-Means only on Devign, a balanced dataset, and observed that both acquisition functions consistently outperformed random selection. However, this dataset carries a certain limitation, i.e., does not reflect the real-world distribution of vulnerabilities, where non-vulnerable samples significantly outnumber vulnerable ones. To better evaluate active learning in a more realistic scenario, we expand the experiments on BigVul and DiverseVul, the imbalanced datasets.
We find that standard active learning strategies do not always outperform random selection in these more practical scenarios. This result suggests that conventional acquisition functions do not always prioritize samples well.

Differently, the active learning enhanced by our proposed approach can produce not only better but also consistent performance.
Additionally, we observe that as active learning progresses through multiple iterations, both standard DeepGini and K-Means exhibit performance fluctuations. This instability arises due to the nature of their selection mechanisms—DeepGini's reliance on uncertainty-based sampling and K-Means’ clustering-based approach introducing variability in data acquisition. We quantify this fluctuation by computing the standard deviation of performance across all experiments in previous research questions, covering three datasets and three models. The results indicate that standard DeepGini and K-Means have significantly higher variance (0.0475 and 0.0685, respectively) compared to our proposed method (0.0405 for DeepGini and 0.0328 for K-Means), demonstrating improved stability. Specifically, our approach reduces variance by 14.7\% for DeepGini and 52.1\% for K-Means, ensuring more consistent training performance across iterations.

Our method, despite some fluctuations, demonstrates reduced variance compared to the standard versions, indicating that \textbf{filtering hard-to-learn data can lead to a more stable training process}. Furthermore, our proposed approach consistently outperforms both standard methods and random selection in the later stages of active learning, reinforcing the benefit of incorporating dataset maps into acquisition functions. These results suggest that our proposed method not only enhances active learning's effectiveness but also mitigates the risk of selecting hard-to-learn data, ultimately leading to improved model performance in vulnerability detection.

\subsection{What Makes a Common Hard-to-learn Instances in Vulnerability Detection?}
Based on our results of RQ2 and RQ3 on Big-Vul \textit{S1}, we observe that CodeBERT, GraphCodeBERT, and CodeT5 identify 207, 197, and 45 samples, respectively, as hard-to-learn data. For the other two datasets, Devign and DiverseVul, CodeBERT identifies 148 and 397 hard-to-learn samples. To better understand the nature of these samples, we analyze their distribution and compare them with their respective \textit{S1} datasets. Through this analysis, we identify the following findings on hard-to-learn instances in vulnerability detection datasets.

\begin{itemize}[wide=0pt]
    \item \textbf{Hard-to-learn instances are more commonly identified in non-vulnerable samples even if the training data is balanced}.
    To address the challenge of the highly imbalanced problem nature, we follow prior work~\cite{yu2019improving} to construct a balanced dataset for model training. In other words, the training data of \textit{S1} is constructed with an equal number of vulnerable and non-vulnerable samples, i.e., 50\% for each category. However, revealed by training dynamics, we find that hard-to-learn data are more commonly found in non-vulnerable samples. More specifically, among the hard-to-learn samples identified by CodeBERT, GraphCodeBERT, and CodeT5 in Big-Vul, non-vulnerable samples account for 79.2\%, 76.1\%, and 80.0\%, respectively. To further validate this trend, we examine the proportion of non-vulnerable samples among the identified hard-to-learn samples in Devign and DiverseVul, which are 61.5\% and 84.4\%, respectively.
    
    \textbf{Our findings contradict the common intuition that models struggle more with learning features from vulnerable samples.}
    Moreover, current vulnerability detection datasets ~\cite{fan2020ac,chakraborty2021deep,nikitopoulos2021crossvul,bhandari2021cvefixes}, are constructed by extracting vulnerable functions from security-fixing commits and non-vulnerable functions from code snapshots taken when no known vulnerabilities were reported. However, this approach may mislabel undetected vulnerabilities as non-vulnerable, leading to misleading training and performance evaluation, which could explain the higher prevalence of hard-to-learn data among non-vulnerable samples.
    Instead, our results suggest that \textbf{more attention is needed on examining non-vulnerable samples in vulnerability datasets}. For instance, examine whether the inaccurate label could be the potential reason making non-vulnerable samples harder for models to learn effectively. Based on our findings, we highlight the importance of model's learning ability on the samples of both sides and advocate to develop more solutions to examine non-vulnerable samples when constructing vulnerability detection datasets.

    \item \textbf{Surprisingly, hard-to-learn instances exist in vulnerable code tend to be shorter, can be patched with a smaller code change, and are also less complex than the average.}

    Intuitively, the more complex the code, the harder to learn. Therefore, we hypothesize that hard-to-learn data are often appear to be complex than average.
    To verify this, we use three indicators to measure the complexity of vulnerability data, (1) code token length, (2) patch size, and (3) Cyclomatic Complexity.

    \begin{enumerate}[wide=0pt]
        \item \textbf{Code length} is often considered as an indicator of code complexity~\cite{munoz2020empirical,klima2022selected,peitek2021program}.
        Considering the training paradigm of LLM, especially tokenization, we particularly focus on token length analysis.
        We aggregate all hard-to-learn samples identified by CodeBERT, GraphCodeBERT, and CodeT5 in Big-Vul and extract 81 vulnerable functions to compare with the complete set of vulnerable functions in \textit{S1}. We find that the average token length of vulnerable functions in \textit{S1} is 531, whereas, surprisingly, in the hard-to-learn set, the average token length is only 104 which is significantly lower. We also perform the same analysis on DiverseVul and Devign. We find that, for DiverseVul, the average token length of vulnerable functions is 766, whereas in hard-to-learn set, it also dramatically lower (89\% relatively) to only 82.7. However, Devign is the exception, we find that the difference is minimal, with the original and hard-to-learn functions having token lengths of 504 and 478, respectively.    
        
        \item \textbf{Patch size} can also be considered as another indicator of complexity in vulnerability context. We hypothesize that, hard-to-learn data tend to be the vulnerable code which require bigger changes to be patched. Fortunately, Big-Vul provides patch information which enables us to compare function modifications before and after patching. Note that, Devign and DiverseVul do not provide patch information. We calculate that, the average patch size (sum of added and deleted lines) in the original \textit{S1} is 9.6, while the patch size of hard-to-learn samples is only 5.65 (41\% smaller).
        
        \item \textbf{Cyclomatic Complexity} has also been used analyze the complexity of the code samples. Specifically, we utlize a widely-used tool named Lizard~\cite{lizard} to compute the complexity of C/C++ functions. In Big-Vul, the average complexity of vulnerable functions in hard-to-learn data is 3.58, compared to 14.8 in the full \textit{S1}. A similar trend is observed in DiverseVul, where the original vulnerable functions have an average complexity of 22.92, whereas the hard-to-learn data have a much lower complexity of 2.43. In contrast, again, Devign does not exhibit this trend, with complexity values for the original and hard-to-learn functions being 11.95 and 11.76, respectively. This suggests that in more realistic datasets, vulnerable functions with lower complexity are more likely to be classified as hard-to-learn, possibly because they lack structural patterns that models rely on for learning vulnerability detection.
    \end{enumerate}
\end{itemize}

In summary, our find that, our analysis on Big-Vul and DiverseVul datasets are anti-intuitive, i.e., \textit{the more complex, the harder to learn}. Instead, it indicates the opposite conclusion, i.e., \textbf{hard-to-learn vulnerable code samples seems more tend to be ``simpler'' than the average}.
Our observation reveals a specific focus that vulnerability detection studies often overlook.
Moreover, we also observe that the Devign dataset is often the exception.
The potential reason could be that, in more realistic (i.e., highly imbalanced) datasets like Big-Vul and DiverseVul, functions with smaller token sizes and smaller patches are more likely to be classified as hard-to-learn due to potential reasons, such as the lack of sufficient context.

\subsection{Threats to Validity}
One threat to internal validity is that, in RQ2, our proposed method modifies the standard active learning pipeline by selecting 125 candidates per iteration instead of 100, then removing the top 20\% most similar to previously identified hard-to-learn data. The 20\% filtering ratio is determined as a reasonable balance between preserving valuable data and removing misleading samples. However, we have not exhaustively tested with alternative ratios. While our experimental setup is consistent, further research is necessary to investigate how different filtering thresholds affect active learning performance. Another threat lies in the parameter setting for identifying hard-to-learn data in RQ1. We enumerate different values of confidence and variability thresholds and based on the empirical results on the validation data, we set confidence to 0.3 and variability to 0.4. To mitigate this threat, we plan to develop an effective and also automated way to identify the optimal values of these thresholds in our future work.

Regarding external validity, one is about the generalizability of our results. In our RQ3, we have specifically investigated this threat by experimenting with more models and datasets. This minimizes concerns about dataset and model dependency.
Another threat is that we only evaluate C/C++ vulnerability datasets, and its generalizability to other languages remains uncertain. Future work will extend the study to other languages and incorporate additional metrics for a more comprehensive evaluation.

\section{Related Works}

\subsection{LLM for Vulnerability Detection}
Recent work in vulnerability detection has increasingly focused on fine-tuning Large Language Models (LLMs) on large, specialized datasets, leveraging their powerful semantic representation capabilities~\cite{akuthota2023vulnerability,chakraborty2024revisiting,ding2024vulnerability,zhou2024large,shestov2025finetuning}. The prevailing assumption has been that larger fine-tuning datasets directly translate to better model performance. However, emerging research challenges this linear relationship. For instance, Chen et al.~\cite{chen2023diversevul} demonstrated that simply increasing the volume of training data does not guarantee improved performance for deep learning models in this domain. This finding suggests that the utility of additional data diminishes, and the composition of the dataset is as important as its size. Compounding this issue is the significant practical barrier to creating large, high-quality datasets: the manual annotation of vulnerabilities is an extremely time-consuming and expensive process that requires deep security
expertise~\cite{ding2024vulnerability,chakraborty2024revisiting,akuthota2023vulnerability,chakraborty2021deep}.

This creates a tough challenge: although large language models are highly capable, blindly expanding labeled datasets is neither cost-effective nor practical. Consequently, the research focus is beginning to shift from simply collecting *more* data towards developing methods that can more efficiently identify *high-value* data. This has spurred interest in techniques that can pinpoint subsets of data offering the highest learning value, thereby maximizing model improvement while respecting the practical constraints of expert annotation. This focus on intelligent and efficient data selection directly motivates our work, which aims to advance these methods by incorporating a deeper understanding of model learning dynamics.

\subsection{Active Learning for Vulnerability Detection}
Active learning has been applied in various software engineering tasks to reduce labeling costs and improve model efficiency. Moskovitch et al.~\cite{moskovitch2008malicious} proposed an early framework utilizing active learning for detecting malicious code. Yu et al.~\cite{yu2019improving} introduced HARMLESS, an active learning-based vulnerability inspection tool that incrementally refines a support vector machine model to guide security experts toward high-risk code regions, effectively reducing the effort needed for vulnerability detection. Lu et al.~\cite{lu2014defect} applied active learning to defect prediction across software versions, combining uncertainty sampling with dimensionality reduction to improve defect detection performance. More recently, Hu et al.~\cite{hu2024active} introduced the first active learning benchmark for code-related tasks, including vulnerability detection, evaluating over ten acquisition functions on pre-trained code models. 

While these works have demonstrated the potential of active learning in vulnerability detection, they have primarily focused on refining acquisition functions based on metrics like model uncertainty. However, only focusing on uncertainty may cause the active learner to prioritize instances that are not efficiently learnable, potentially leading to a less optimal training path. In our work, we address this limitation by integrating dataset maps into active learning.

\section{Conclusions}
In this work, we investigate a key challenge in vulnerability detection: the inefficient use of labeled data for model training. To address this, we propose an approach that enhances active learning with insights from dataset maps. The key idea is to leverage training dynamics to identify hard-to-learn samples, using this information to create a more intelligent selection strategy. Our initial results demonstrate that this learnability-aware approach leads to significant performance gains. Building on this, we explore the potential of optimizing active learning for vulnerability detection. While existing active learning methods primarily focus on refining acquisition functions, our findings confirm that these approaches do not always offer a consistent advantage over random selection. In contrast, our results demonstrate that our proposed method consistently outperforms both standard active learning and random selection. This leads to more stable and significant improvements in model performance across multiple models and datasets.

Looking forward, we plan to develop automated methods to categorize hard/ambiguous/easy-to-learn samples, making our framework more adaptive. Another direction is to analyze the root causes of learning difficulty. This involves studying the complex relationship between the traits of hard-to-learn samples and various data quality issues in the specific vulnerability detection context, such as the ambiguity of non-vulnerable labels, which could help improve future learning strategies and dataset construction.



\bibliographystyle{ACM-Reference-Format}
\bibliography{refs}


\begin{thebibliography}{52}


\ifx \showCODEN    \undefined \def \showCODEN     #1{\unskip}     \fi
\ifx \showISBNx    \undefined \def \showISBNx     #1{\unskip}     \fi
\ifx \showISBNxiii \undefined \def \showISBNxiii  #1{\unskip}     \fi
\ifx \showISSN     \undefined \def \showISSN      #1{\unskip}     \fi
\ifx \showLCCN     \undefined \def \showLCCN      #1{\unskip}     \fi
\ifx \shownote     \undefined \def \shownote      #1{#1}          \fi
\ifx \showarticletitle \undefined \def \showarticletitle #1{#1}   \fi
\ifx \showURL      \undefined \def \showURL       {\relax}        \fi
\providecommand\bibfield[2]{#2}
\providecommand\bibinfo[2]{#2}
\providecommand\natexlab[1]{#1}
\providecommand\showeprint[2][]{arXiv:#2}

\bibitem[Akuthota et~al\mbox{.}(2023)]%
        {akuthota2023vulnerability}
\bibfield{author}{\bibinfo{person}{Vishwanath Akuthota}, \bibinfo{person}{Raghunandan Kasula}, \bibinfo{person}{Sabiha~Tasnim Sumona}, \bibinfo{person}{Masud Mohiuddin}, \bibinfo{person}{Md~Tanzim Reza}, {and} \bibinfo{person}{Md~Mizanur Rahman}.} \bibinfo{year}{2023}\natexlab{}.
\newblock \showarticletitle{Vulnerability detection and monitoring using llm}. In \bibinfo{booktitle}{\emph{2023 IEEE 9th International Women in Engineering (WIE) Conference on Electrical and Computer Engineering (WIECON-ECE)}}. IEEE, \bibinfo{pages}{309--314}.
\newblock


\bibitem[Amasaki(2020)]%
        {amasaki2020cross}
\bibfield{author}{\bibinfo{person}{Sousuke Amasaki}.} \bibinfo{year}{2020}\natexlab{}.
\newblock \showarticletitle{Cross-version defect prediction: use historical data, cross-project data, or both?}
\newblock \bibinfo{journal}{\emph{Empirical Software Engineering}}  \bibinfo{volume}{25} (\bibinfo{year}{2020}), \bibinfo{pages}{1573--1595}.
\newblock


\bibitem[Anonymous(2025)]%
        {badseeds}
\bibfield{author}{\bibinfo{person}{Anonymous}.} \bibinfo{year}{2025}\natexlab{}.
\newblock \bibinfo{title}{Replication Package for "Smart Cuts"}.
\newblock
\urldef\tempurl%
\url{https://anonymous.4open.science/r/test-572B/}
\showURL{%
\tempurl}
\newblock
\shownote{Accessed: March 2025}.


\bibitem[Beluch et~al\mbox{.}(2018)]%
        {beluch2018power}
\bibfield{author}{\bibinfo{person}{William~H Beluch}, \bibinfo{person}{Tim Genewein}, \bibinfo{person}{Andreas N{\"u}rnberger}, {and} \bibinfo{person}{Jan~M K{\"o}hler}.} \bibinfo{year}{2018}\natexlab{}.
\newblock \showarticletitle{The power of ensembles for active learning in image classification}. In \bibinfo{booktitle}{\emph{Proceedings of the IEEE conference on computer vision and pattern recognition}}. \bibinfo{pages}{9368--9377}.
\newblock


\bibitem[Bhandari et~al\mbox{.}(2021)]%
        {bhandari2021cvefixes}
\bibfield{author}{\bibinfo{person}{Guru Bhandari}, \bibinfo{person}{Amara Naseer}, {and} \bibinfo{person}{Leon Moonen}.} \bibinfo{year}{2021}\natexlab{}.
\newblock \showarticletitle{CVEfixes: automated collection of vulnerabilities and their fixes from open-source software}. In \bibinfo{booktitle}{\emph{Proceedings of the 17th International Conference on Predictive Models and Data Analytics in Software Engineering}}. \bibinfo{pages}{30--39}.
\newblock


\bibitem[Bilgic and Getoor(2009)]%
        {bilgic2009link}
\bibfield{author}{\bibinfo{person}{Mustafa Bilgic} {and} \bibinfo{person}{Lise Getoor}.} \bibinfo{year}{2009}\natexlab{}.
\newblock \showarticletitle{Link-based active learning}. In \bibinfo{booktitle}{\emph{NIPS workshop on analyzing networks and learning with graphs}}, Vol.~\bibinfo{volume}{4}. \bibinfo{pages}{9}.
\newblock


\bibitem[Chakraborty et~al\mbox{.}(2024)]%
        {chakraborty2024revisiting}
\bibfield{author}{\bibinfo{person}{Partha Chakraborty}, \bibinfo{person}{Krishna~Kanth Arumugam}, \bibinfo{person}{Mahmoud Alfadel}, \bibinfo{person}{Meiyappan Nagappan}, {and} \bibinfo{person}{Shane McIntosh}.} \bibinfo{year}{2024}\natexlab{}.
\newblock \showarticletitle{Revisiting the performance of deep learning-based vulnerability detection on realistic datasets}.
\newblock \bibinfo{journal}{\emph{IEEE Transactions on Software Engineering}} (\bibinfo{year}{2024}).
\newblock


\bibitem[Chakraborty et~al\mbox{.}(2021)]%
        {chakraborty2021deep}
\bibfield{author}{\bibinfo{person}{Saikat Chakraborty}, \bibinfo{person}{Rahul Krishna}, \bibinfo{person}{Yangruibo Ding}, {and} \bibinfo{person}{Baishakhi Ray}.} \bibinfo{year}{2021}\natexlab{}.
\newblock \showarticletitle{Deep learning based vulnerability detection: Are we there yet?}
\newblock \bibinfo{journal}{\emph{IEEE Transactions on Software Engineering}} \bibinfo{volume}{48}, \bibinfo{number}{9} (\bibinfo{year}{2021}), \bibinfo{pages}{3280--3296}.
\newblock


\bibitem[Chen et~al\mbox{.}(2023)]%
        {chen2023diversevul}
\bibfield{author}{\bibinfo{person}{Yizheng Chen}, \bibinfo{person}{Zhoujie Ding}, \bibinfo{person}{Lamya Alowain}, \bibinfo{person}{Xinyun Chen}, {and} \bibinfo{person}{David Wagner}.} \bibinfo{year}{2023}\natexlab{}.
\newblock \showarticletitle{Diversevul: A new vulnerable source code dataset for deep learning based vulnerability detection}. In \bibinfo{booktitle}{\emph{Proceedings of the 26th International Symposium on Research in Attacks, Intrusions and Defenses}}. \bibinfo{pages}{654--668}.
\newblock


\bibitem[Ding et~al\mbox{.}(2024)]%
        {ding2024vulnerability}
\bibfield{author}{\bibinfo{person}{Yangruibo Ding}, \bibinfo{person}{Yanjun Fu}, \bibinfo{person}{Omniyyah Ibrahim}, \bibinfo{person}{Chawin Sitawarin}, \bibinfo{person}{Xinyun Chen}, \bibinfo{person}{Basel Alomair}, \bibinfo{person}{David Wagner}, \bibinfo{person}{Baishakhi Ray}, {and} \bibinfo{person}{Yizheng Chen}.} \bibinfo{year}{2024}\natexlab{}.
\newblock \showarticletitle{Vulnerability Detection with Code Language Models: How Far Are We?}. In \bibinfo{booktitle}{\emph{2025 IEEE/ACM 47th International Conference on Software Engineering (ICSE)}}. IEEE Computer Society, \bibinfo{pages}{469--481}.
\newblock


\bibitem[Du et~al\mbox{.}(2024)]%
        {du2024generalization}
\bibfield{author}{\bibinfo{person}{Xiaohu Du}, \bibinfo{person}{Ming Wen}, \bibinfo{person}{Jiahao Zhu}, \bibinfo{person}{Zifan Xie}, \bibinfo{person}{Bin Ji}, \bibinfo{person}{Huijun Liu}, \bibinfo{person}{Xuanhua Shi}, {and} \bibinfo{person}{Hai Jin}.} \bibinfo{year}{2024}\natexlab{}.
\newblock \showarticletitle{Generalization-Enhanced Code Vulnerability Detection via Multi-Task Instruction Fine-Tuning}. In \bibinfo{booktitle}{\emph{Findings of the Association for Computational Linguistics ACL 2024}}. \bibinfo{pages}{10507--10521}.
\newblock


\bibitem[Fan et~al\mbox{.}(2020)]%
        {fan2020ac}
\bibfield{author}{\bibinfo{person}{Jiahao Fan}, \bibinfo{person}{Yi Li}, \bibinfo{person}{Shaohua Wang}, {and} \bibinfo{person}{Tien~N Nguyen}.} \bibinfo{year}{2020}\natexlab{}.
\newblock \showarticletitle{AC/C++ code vulnerability dataset with code changes and CVE summaries}. In \bibinfo{booktitle}{\emph{Proceedings of the 17th international conference on mining software repositories}}. \bibinfo{pages}{508--512}.
\newblock


\bibitem[Feng et~al\mbox{.}(2020b)]%
        {feng2020deepgini}
\bibfield{author}{\bibinfo{person}{Yang Feng}, \bibinfo{person}{Qingkai Shi}, \bibinfo{person}{Xinyu Gao}, \bibinfo{person}{Jun Wan}, \bibinfo{person}{Chunrong Fang}, {and} \bibinfo{person}{Zhenyu Chen}.} \bibinfo{year}{2020}\natexlab{b}.
\newblock \showarticletitle{Deepgini: prioritizing massive tests to enhance the robustness of deep neural networks}. In \bibinfo{booktitle}{\emph{Proceedings of the 29th ACM SIGSOFT international symposium on software testing and analysis}}. \bibinfo{pages}{177--188}.
\newblock


\bibitem[Feng et~al\mbox{.}(2020a)]%
        {feng2020codebert}
\bibfield{author}{\bibinfo{person}{Zhangyin Feng}, \bibinfo{person}{Daya Guo}, \bibinfo{person}{Duyu Tang}, \bibinfo{person}{Nan Duan}, \bibinfo{person}{Xiaocheng Feng}, \bibinfo{person}{Ming Gong}, \bibinfo{person}{Linjun Shou}, \bibinfo{person}{Bing Qin}, \bibinfo{person}{Ting Liu}, \bibinfo{person}{Daxin Jiang}, {et~al\mbox{.}}} \bibinfo{year}{2020}\natexlab{a}.
\newblock \showarticletitle{CodeBERT: A Pre-Trained Model for Programming and Natural Languages}. In \bibinfo{booktitle}{\emph{Findings of the Association for Computational Linguistics: EMNLP 2020}}. \bibinfo{pages}{1536--1547}.
\newblock


\bibitem[Fu and Tantithamthavorn(2022)]%
        {fu2022linevul}
\bibfield{author}{\bibinfo{person}{Michael Fu} {and} \bibinfo{person}{Chakkrit Tantithamthavorn}.} \bibinfo{year}{2022}\natexlab{}.
\newblock \showarticletitle{Linevul: A transformer-based line-level vulnerability prediction}. In \bibinfo{booktitle}{\emph{Proceedings of the 19th International Conference on Mining Software Repositories}}. \bibinfo{pages}{608--620}.
\newblock


\bibitem[Gal et~al\mbox{.}(2017)]%
        {gal2017deep}
\bibfield{author}{\bibinfo{person}{Yarin Gal}, \bibinfo{person}{Riashat Islam}, {and} \bibinfo{person}{Zoubin Ghahramani}.} \bibinfo{year}{2017}\natexlab{}.
\newblock \showarticletitle{Deep bayesian active learning with image data}. In \bibinfo{booktitle}{\emph{International conference on machine learning}}. PMLR, \bibinfo{pages}{1183--1192}.
\newblock


\bibitem[Guo et~al\mbox{.}({[n.\,d.]})]%
        {guographcodebert}
\bibfield{author}{\bibinfo{person}{Daya Guo}, \bibinfo{person}{Shuo Ren}, \bibinfo{person}{Shuai Lu}, \bibinfo{person}{Zhangyin Feng}, \bibinfo{person}{Duyu Tang}, \bibinfo{person}{LIU Shujie}, \bibinfo{person}{Long Zhou}, \bibinfo{person}{Nan Duan}, \bibinfo{person}{Alexey Svyatkovskiy}, \bibinfo{person}{Shengyu Fu}, {et~al\mbox{.}}} \bibinfo{year}{[n.\,d.]}\natexlab{}.
\newblock \showarticletitle{GraphCodeBERT: Pre-training Code Representations with Data Flow}. In \bibinfo{booktitle}{\emph{International Conference on Learning Representations}}.
\newblock


\bibitem[Guo(2010)]%
        {guo2010active}
\bibfield{author}{\bibinfo{person}{Yuhong Guo}.} \bibinfo{year}{2010}\natexlab{}.
\newblock \showarticletitle{Active instance sampling via matrix partition}.
\newblock \bibinfo{journal}{\emph{Advances in neural information processing systems}}  \bibinfo{volume}{23} (\bibinfo{year}{2010}).
\newblock


\bibitem[Hu et~al\mbox{.}(2021)]%
        {hu2021towards}
\bibfield{author}{\bibinfo{person}{Qiang Hu}, \bibinfo{person}{Yuejun Guo}, \bibinfo{person}{Maxime Cordy}, \bibinfo{person}{Xiaofei Xie}, \bibinfo{person}{Wei Ma}, \bibinfo{person}{Mike Papadakis}, {and} \bibinfo{person}{Yves Le~Traon}.} \bibinfo{year}{2021}\natexlab{}.
\newblock \showarticletitle{Towards exploring the limitations of active learning: An empirical study}. In \bibinfo{booktitle}{\emph{2021 36th IEEE/ACM International Conference on Automated Software Engineering (ASE)}}. IEEE, \bibinfo{pages}{917--929}.
\newblock


\bibitem[Hu et~al\mbox{.}(2024)]%
        {hu2024active}
\bibfield{author}{\bibinfo{person}{Qiang Hu}, \bibinfo{person}{Yuejun Guo}, \bibinfo{person}{Xiaofei Xie}, \bibinfo{person}{Maxime Cordy}, \bibinfo{person}{Lei Ma}, \bibinfo{person}{Mike Papadakis}, {and} \bibinfo{person}{Yves Le~Traon}.} \bibinfo{year}{2024}\natexlab{}.
\newblock \showarticletitle{Active code learning: Benchmarking sample-efficient training of code models}.
\newblock \bibinfo{journal}{\emph{IEEE Transactions on Software Engineering}} (\bibinfo{year}{2024}).
\newblock


\bibitem[Joshi et~al\mbox{.}(2009)]%
        {joshi2009multi}
\bibfield{author}{\bibinfo{person}{Ajay~J Joshi}, \bibinfo{person}{Fatih Porikli}, {and} \bibinfo{person}{Nikolaos Papanikolopoulos}.} \bibinfo{year}{2009}\natexlab{}.
\newblock \showarticletitle{Multi-class active learning for image classification}. In \bibinfo{booktitle}{\emph{2009 ieee conference on computer vision and pattern recognition}}. IEEE, \bibinfo{pages}{2372--2379}.
\newblock


\bibitem[Klima et~al\mbox{.}(2022)]%
        {klima2022selected}
\bibfield{author}{\bibinfo{person}{Matej Klima}, \bibinfo{person}{Miroslav Bures}, \bibinfo{person}{Karel Frajtak}, \bibinfo{person}{Vaclav Rechtberger}, \bibinfo{person}{Michal Trnka}, \bibinfo{person}{Xavier Bellekens}, \bibinfo{person}{Tomas Cerny}, {and} \bibinfo{person}{Bestoun~S Ahmed}.} \bibinfo{year}{2022}\natexlab{}.
\newblock \showarticletitle{Selected code-quality characteristics and metrics for internet of things systems}.
\newblock \bibinfo{journal}{\emph{IEEE Access}}  \bibinfo{volume}{10} (\bibinfo{year}{2022}), \bibinfo{pages}{46144--46161}.
\newblock


\bibitem[Li et~al\mbox{.}(2012)]%
        {li2012sample}
\bibfield{author}{\bibinfo{person}{Ming Li}, \bibinfo{person}{Hongyu Zhang}, \bibinfo{person}{Rongxin Wu}, {and} \bibinfo{person}{Zhi-Hua Zhou}.} \bibinfo{year}{2012}\natexlab{}.
\newblock \showarticletitle{Sample-based software defect prediction with active and semi-supervised learning}.
\newblock \bibinfo{journal}{\emph{Automated Software Engineering}}  \bibinfo{volume}{19} (\bibinfo{year}{2012}), \bibinfo{pages}{201--230}.
\newblock


\bibitem[Li and Guo(2013)]%
        {li2013adaptive}
\bibfield{author}{\bibinfo{person}{Xin Li} {and} \bibinfo{person}{Yuhong Guo}.} \bibinfo{year}{2013}\natexlab{}.
\newblock \showarticletitle{Adaptive active learning for image classification}. In \bibinfo{booktitle}{\emph{Proceedings of the IEEE conference on computer vision and pattern recognition}}. \bibinfo{pages}{859--866}.
\newblock


\bibitem[Lin et~al\mbox{.}(2020)]%
        {lin2020software}
\bibfield{author}{\bibinfo{person}{Guanjun Lin}, \bibinfo{person}{Sheng Wen}, \bibinfo{person}{Qing-Long Han}, \bibinfo{person}{Jun Zhang}, {and} \bibinfo{person}{Yang Xiang}.} \bibinfo{year}{2020}\natexlab{}.
\newblock \showarticletitle{Software vulnerability detection using deep neural networks: a survey}.
\newblock \bibinfo{journal}{\emph{Proc. IEEE}} \bibinfo{volume}{108}, \bibinfo{number}{10} (\bibinfo{year}{2020}), \bibinfo{pages}{1825--1848}.
\newblock


\bibitem[Liu et~al\mbox{.}(2024)]%
        {liu2024pre}
\bibfield{author}{\bibinfo{person}{Zhongxin Liu}, \bibinfo{person}{Zhijie Tang}, \bibinfo{person}{Junwei Zhang}, \bibinfo{person}{Xin Xia}, {and} \bibinfo{person}{Xiaohu Yang}.} \bibinfo{year}{2024}\natexlab{}.
\newblock \showarticletitle{Pre-training by predicting program dependencies for vulnerability analysis tasks}. In \bibinfo{booktitle}{\emph{Proceedings of the IEEE/ACM 46th International Conference on Software Engineering}}. \bibinfo{pages}{1--13}.
\newblock


\bibitem[Lu et~al\mbox{.}(2014)]%
        {lu2014defect}
\bibfield{author}{\bibinfo{person}{Huihua Lu}, \bibinfo{person}{Ekrem Kocaguneli}, {and} \bibinfo{person}{Bojan Cukic}.} \bibinfo{year}{2014}\natexlab{}.
\newblock \showarticletitle{Defect prediction between software versions with active learning and dimensionality reduction}. In \bibinfo{booktitle}{\emph{2014 IEEE 25th international symposium on software reliability engineering}}. IEEE, \bibinfo{pages}{312--322}.
\newblock


\bibitem[McIntosh and Kamei(2018)]%
        {mcintosh2018fix}
\bibfield{author}{\bibinfo{person}{Shane McIntosh} {and} \bibinfo{person}{Yasutaka Kamei}.} \bibinfo{year}{2018}\natexlab{}.
\newblock \showarticletitle{Are fix-inducing changes a moving target? a longitudinal case study of just-in-time defect prediction}. In \bibinfo{booktitle}{\emph{Proceedings of the 40th international conference on software engineering}}. \bibinfo{pages}{560--560}.
\newblock


\bibitem[Moskovitch et~al\mbox{.}(2008)]%
        {moskovitch2008malicious}
\bibfield{author}{\bibinfo{person}{Robert Moskovitch}, \bibinfo{person}{Nir Nissim}, {and} \bibinfo{person}{Yuval Elovici}.} \bibinfo{year}{2008}\natexlab{}.
\newblock \showarticletitle{Malicious code detection using active learning}. In \bibinfo{booktitle}{\emph{International Workshop on Privacy, Security, and Trust in KDD}}. Springer, \bibinfo{pages}{74--91}.
\newblock


\bibitem[Mu{\~n}oz~Bar{\'o}n et~al\mbox{.}(2020)]%
        {munoz2020empirical}
\bibfield{author}{\bibinfo{person}{Marvin Mu{\~n}oz~Bar{\'o}n}, \bibinfo{person}{Marvin Wyrich}, {and} \bibinfo{person}{Stefan Wagner}.} \bibinfo{year}{2020}\natexlab{}.
\newblock \showarticletitle{An empirical validation of cognitive complexity as a measure of source code understandability}. In \bibinfo{booktitle}{\emph{Proceedings of the 14th ACM/IEEE international symposium on empirical software engineering and measurement (ESEM)}}. \bibinfo{pages}{1--12}.
\newblock


\bibitem[Nikitopoulos et~al\mbox{.}(2021)]%
        {nikitopoulos2021crossvul}
\bibfield{author}{\bibinfo{person}{Georgios Nikitopoulos}, \bibinfo{person}{Konstantina Dritsa}, \bibinfo{person}{Panos Louridas}, {and} \bibinfo{person}{Dimitris Mitropoulos}.} \bibinfo{year}{2021}\natexlab{}.
\newblock \showarticletitle{CrossVul: a cross-language vulnerability dataset with commit data}. In \bibinfo{booktitle}{\emph{Proceedings of the 29th ACM Joint Meeting on European Software Engineering Conference and Symposium on the Foundations of Software Engineering}}. \bibinfo{pages}{1565--1569}.
\newblock


\bibitem[Owen(1965)]%
        {owen1965power}
\bibfield{author}{\bibinfo{person}{Donald~B Owen}.} \bibinfo{year}{1965}\natexlab{}.
\newblock \showarticletitle{The power of Student's t-test}.
\newblock \bibinfo{journal}{\emph{J. Amer. Statist. Assoc.}} \bibinfo{volume}{60}, \bibinfo{number}{309} (\bibinfo{year}{1965}), \bibinfo{pages}{320--333}.
\newblock


\bibitem[Peitek et~al\mbox{.}(2021)]%
        {peitek2021program}
\bibfield{author}{\bibinfo{person}{Norman Peitek}, \bibinfo{person}{Sven Apel}, \bibinfo{person}{Chris Parnin}, \bibinfo{person}{Andr{\'e} Brechmann}, {and} \bibinfo{person}{Janet Siegmund}.} \bibinfo{year}{2021}\natexlab{}.
\newblock \showarticletitle{Program comprehension and code complexity metrics: An fmri study}. In \bibinfo{booktitle}{\emph{2021 IEEE/ACM 43rd International Conference on Software Engineering (ICSE)}}. IEEE, \bibinfo{pages}{524--536}.
\newblock


\bibitem[Raffel et~al\mbox{.}(2020)]%
        {raffel2020exploring}
\bibfield{author}{\bibinfo{person}{Colin Raffel}, \bibinfo{person}{Noam Shazeer}, \bibinfo{person}{Adam Roberts}, \bibinfo{person}{Katherine Lee}, \bibinfo{person}{Sharan Narang}, \bibinfo{person}{Michael Matena}, \bibinfo{person}{Yanqi Zhou}, \bibinfo{person}{Wei Li}, {and} \bibinfo{person}{Peter~J Liu}.} \bibinfo{year}{2020}\natexlab{}.
\newblock \showarticletitle{Exploring the limits of transfer learning with a unified text-to-text transformer}.
\newblock \bibinfo{journal}{\emph{Journal of machine learning research}} \bibinfo{volume}{21}, \bibinfo{number}{140} (\bibinfo{year}{2020}), \bibinfo{pages}{1--67}.
\newblock


\bibitem[Rahman et~al\mbox{.}(2013)]%
        {rahman2013sample}
\bibfield{author}{\bibinfo{person}{Foyzur Rahman}, \bibinfo{person}{Daryl Posnett}, \bibinfo{person}{Israel Herraiz}, {and} \bibinfo{person}{Premkumar Devanbu}.} \bibinfo{year}{2013}\natexlab{}.
\newblock \showarticletitle{Sample size vs. bias in defect prediction}. In \bibinfo{booktitle}{\emph{Proceedings of the 2013 9th joint meeting on foundations of software engineering}}. \bibinfo{pages}{147--157}.
\newblock


\bibitem[Sener and Savarese(2018)]%
        {sener2018active}
\bibfield{author}{\bibinfo{person}{Ozan Sener} {and} \bibinfo{person}{Silvio Savarese}.} \bibinfo{year}{2018}\natexlab{}.
\newblock \showarticletitle{Active Learning for Convolutional Neural Networks: A Core-Set Approach}. In \bibinfo{booktitle}{\emph{International Conference on Learning Representations}}.
\newblock


\bibitem[Settles(2009)]%
        {settles2009active}
\bibfield{author}{\bibinfo{person}{Burr Settles}.} \bibinfo{year}{2009}\natexlab{}.
\newblock \showarticletitle{Active learning literature survey}.
\newblock  (\bibinfo{year}{2009}).
\newblock


\bibitem[Shestov et~al\mbox{.}(2025)]%
        {shestov2025finetuning}
\bibfield{author}{\bibinfo{person}{Aleksei Shestov}, \bibinfo{person}{Rodion Levichev}, \bibinfo{person}{Ravil Mussabayev}, \bibinfo{person}{Evgeny Maslov}, \bibinfo{person}{Pavel Zadorozhny}, \bibinfo{person}{Anton Cheshkov}, \bibinfo{person}{Rustam Mussabayev}, \bibinfo{person}{Alymzhan Toleu}, \bibinfo{person}{Gulmira Tolegen}, {and} \bibinfo{person}{Alexander Krassovitskiy}.} \bibinfo{year}{2025}\natexlab{}.
\newblock \showarticletitle{Finetuning large language models for vulnerability detection}.
\newblock \bibinfo{journal}{\emph{IEEE Access}} (\bibinfo{year}{2025}).
\newblock


\bibitem[Steenhoek et~al\mbox{.}(2024)]%
        {steenhoek2024dataflow}
\bibfield{author}{\bibinfo{person}{Benjamin Steenhoek}, \bibinfo{person}{Hongyang Gao}, {and} \bibinfo{person}{Wei Le}.} \bibinfo{year}{2024}\natexlab{}.
\newblock \showarticletitle{Dataflow analysis-inspired deep learning for efficient vulnerability detection}. In \bibinfo{booktitle}{\emph{Proceedings of the 46th ieee/acm international conference on software engineering}}. \bibinfo{pages}{1--13}.
\newblock


\bibitem[Steenhoek et~al\mbox{.}(2023)]%
        {steenhoek2023empirical}
\bibfield{author}{\bibinfo{person}{Benjamin Steenhoek}, \bibinfo{person}{Md~Mahbubur Rahman}, \bibinfo{person}{Richard Jiles}, {and} \bibinfo{person}{Wei Le}.} \bibinfo{year}{2023}\natexlab{}.
\newblock \showarticletitle{An empirical study of deep learning models for vulnerability detection}. In \bibinfo{booktitle}{\emph{2023 IEEE/ACM 45th International Conference on Software Engineering (ICSE)}}. IEEE, \bibinfo{pages}{2237--2248}.
\newblock


\bibitem[Swayamdipta et~al\mbox{.}(2020)]%
        {swayamdipta2020dataset}
\bibfield{author}{\bibinfo{person}{Swabha Swayamdipta}, \bibinfo{person}{Roy Schwartz}, \bibinfo{person}{Nicholas Lourie}, \bibinfo{person}{Yizhong Wang}, \bibinfo{person}{Hannaneh Hajishirzi}, \bibinfo{person}{Noah~A Smith}, {and} \bibinfo{person}{Yejin Choi}.} \bibinfo{year}{2020}\natexlab{}.
\newblock \showarticletitle{Dataset Cartography: Mapping and Diagnosing Datasets with Training Dynamics}. In \bibinfo{booktitle}{\emph{Proceedings of the 2020 Conference on Empirical Methods in Natural Language Processing (EMNLP)}}. \bibinfo{pages}{9275--9293}.
\newblock


\bibitem[Tong and Koller(2001)]%
        {tong2001support}
\bibfield{author}{\bibinfo{person}{Simon Tong} {and} \bibinfo{person}{Daphne Koller}.} \bibinfo{year}{2001}\natexlab{}.
\newblock \showarticletitle{Support vector machine active learning with applications to text classification}.
\newblock \bibinfo{journal}{\emph{Journal of machine learning research}} \bibinfo{volume}{2}, \bibinfo{number}{Nov} (\bibinfo{year}{2001}), \bibinfo{pages}{45--66}.
\newblock


\bibitem[Wang et~al\mbox{.}(2021)]%
        {wang2021codet5}
\bibfield{author}{\bibinfo{person}{Yue Wang}, \bibinfo{person}{Weishi Wang}, \bibinfo{person}{Shafiq Joty}, {and} \bibinfo{person}{Steven~CH Hoi}.} \bibinfo{year}{2021}\natexlab{}.
\newblock \showarticletitle{CodeT5: Identifier-aware Unified Pre-trained Encoder-Decoder Models for Code Understanding and Generation}. In \bibinfo{booktitle}{\emph{Proceedings of the 2021 Conference on Empirical Methods in Natural Language Processing}}. \bibinfo{pages}{8696--8708}.
\newblock


\bibitem[Wu et~al\mbox{.}(2022)]%
        {wu2022vulcnn}
\bibfield{author}{\bibinfo{person}{Yueming Wu}, \bibinfo{person}{Deqing Zou}, \bibinfo{person}{Shihan Dou}, \bibinfo{person}{Wei Yang}, \bibinfo{person}{Duo Xu}, {and} \bibinfo{person}{Hai Jin}.} \bibinfo{year}{2022}\natexlab{}.
\newblock \showarticletitle{Vulcnn: An image-inspired scalable vulnerability detection system}. In \bibinfo{booktitle}{\emph{Proceedings of the 44th International Conference on Software Engineering}}. \bibinfo{pages}{2365--2376}.
\newblock


\bibitem[Xu et~al\mbox{.}(2018)]%
        {xu2018cross}
\bibfield{author}{\bibinfo{person}{Zhou Xu}, \bibinfo{person}{Jin Liu}, \bibinfo{person}{Xiapu Luo}, {and} \bibinfo{person}{Tao Zhang}.} \bibinfo{year}{2018}\natexlab{}.
\newblock \showarticletitle{Cross-version defect prediction via hybrid active learning with kernel principal component analysis}. In \bibinfo{booktitle}{\emph{2018 IEEE 25th international conference on software analysis, evolution and reengineering (SANER)}}. IEEE, \bibinfo{pages}{209--220}.
\newblock


\bibitem[Yang et~al\mbox{.}(2003)]%
        {yang2003automatically}
\bibfield{author}{\bibinfo{person}{Jie Yang} {et~al\mbox{.}}} \bibinfo{year}{2003}\natexlab{}.
\newblock \showarticletitle{Automatically labeling video data using multi-class active learning}. In \bibinfo{booktitle}{\emph{Proceedings Ninth IEEE international conference on computer vision}}. IEEE, \bibinfo{pages}{516--523}.
\newblock


\bibitem[Yin(2025)]%
        {lizard}
\bibfield{author}{\bibinfo{person}{Terry Yin}.} \bibinfo{year}{2025}\natexlab{}.
\newblock \bibinfo{title}{Lizard - An Extensible Cyclomatic Complexity Analyzer}.
\newblock \bibinfo{howpublished}{\url{https://pypi.org/project/lizard/}}.
\newblock


\bibitem[Yu et~al\mbox{.}(2019)]%
        {yu2019improving}
\bibfield{author}{\bibinfo{person}{Zhe Yu}, \bibinfo{person}{Christopher Theisen}, \bibinfo{person}{Laurie Williams}, {and} \bibinfo{person}{Tim Menzies}.} \bibinfo{year}{2019}\natexlab{}.
\newblock \showarticletitle{Improving vulnerability inspection efficiency using active learning}.
\newblock \bibinfo{journal}{\emph{IEEE Transactions on Software Engineering}} \bibinfo{volume}{47}, \bibinfo{number}{11} (\bibinfo{year}{2019}), \bibinfo{pages}{2401--2420}.
\newblock


\bibitem[Zhang et~al\mbox{.}(2022)]%
        {zhang2022survey}
\bibfield{author}{\bibinfo{person}{Zhisong Zhang}, \bibinfo{person}{Emma Strubell}, {and} \bibinfo{person}{Eduard Hovy}.} \bibinfo{year}{2022}\natexlab{}.
\newblock \showarticletitle{A Survey of Active Learning for Natural Language Processing}. In \bibinfo{booktitle}{\emph{Proceedings of the 2022 Conference on Empirical Methods in Natural Language Processing}}. \bibinfo{pages}{6166--6190}.
\newblock


\bibitem[Zhao et~al\mbox{.}(2024)]%
        {zhao2024coding}
\bibfield{author}{\bibinfo{person}{Yu Zhao}, \bibinfo{person}{Lina Gong}, \bibinfo{person}{Zhiqiu Huang}, \bibinfo{person}{Yongwei Wang}, \bibinfo{person}{Mingqiang Wei}, {and} \bibinfo{person}{Fei Wu}.} \bibinfo{year}{2024}\natexlab{}.
\newblock \showarticletitle{Coding-ptms: How to find optimal code pre-trained models for code embedding in vulnerability detection?}. In \bibinfo{booktitle}{\emph{Proceedings of the 39th IEEE/ACM International Conference on Automated Software Engineering}}. \bibinfo{pages}{1732--1744}.
\newblock


\bibitem[Zhou et~al\mbox{.}(2024)]%
        {zhou2024large}
\bibfield{author}{\bibinfo{person}{Xin Zhou}, \bibinfo{person}{Sicong Cao}, \bibinfo{person}{Xiaobing Sun}, {and} \bibinfo{person}{David Lo}.} \bibinfo{year}{2024}\natexlab{}.
\newblock \showarticletitle{Large language model for vulnerability detection and repair: Literature review and the road ahead}.
\newblock \bibinfo{journal}{\emph{ACM Transactions on Software Engineering and Methodology}} (\bibinfo{year}{2024}).
\newblock


\bibitem[Zhou et~al\mbox{.}(2019)]%
        {zhou2019devign}
\bibfield{author}{\bibinfo{person}{Yaqin Zhou}, \bibinfo{person}{Shangqing Liu}, \bibinfo{person}{Jingkai Siow}, \bibinfo{person}{Xiaoning Du}, {and} \bibinfo{person}{Yang Liu}.} \bibinfo{year}{2019}\natexlab{}.
\newblock \showarticletitle{Devign: Effective vulnerability identification by learning comprehensive program semantics via graph neural networks}.
\newblock \bibinfo{journal}{\emph{Advances in neural information processing systems}}  \bibinfo{volume}{32} (\bibinfo{year}{2019}).
\newblock


\end{thebibliography}

\end{document}